\documentclass[aps,pra,10pt,twocolumn,notitlepage,superscriptaddress,dvipsnames,longbibliography,tightenlines,floatfix]{revtex4-2}
\usepackage[T1]{fontenc}
\usepackage[table]{xcolor}
\usepackage{bbold}
\usepackage{mathtools}
\usepackage{bibunits}
\usepackage{siunitx}
\usepackage[toc,page,title]{appendix}
\usepackage{graphicx}
\usepackage[caption=false, subrefformat=parens, labelformat=parens]{subfig}
\graphicspath{{./circuits/}, {./figures/}}

\usepackage{xcolor}
\definecolor{dark-red}{rgb}{0.4,0.15,0.15}
\definecolor{dark-blue}{rgb}{0.15,0.15,0.4}
\definecolor{medium-blue}{rgb}{0,0,0.5}
\usepackage[hypertexnames=false]{hyperref}
\usepackage[all]{hypcap}
\usepackage{cleveref}
\hypersetup{
    colorlinks, linkcolor={dark-red},
    citecolor={dark-blue}, urlcolor={medium-blue}
}

\newcommand{\nsp}{\hspace{-0.4pt}}
\newcommand{\xssp}{\hspace{0.4pt}}
\newcommand{\norm}[1]{\lvert #1 \rvert}
\newcommand{\opnorm}[1]{\lVert #1 \rVert}
\newcommand{\normb}[1]{\big\lvert #1 \big\rvert}

\newcommand{\ket}[1]{\lvert\, #1\, \rangle}

\newcommand{\bra}[1]{\langle\, #1\, \rvert}

\newcommand{\braket}[2]{\langle\, #1\,\vert\, #2 \,\rangle}

\newcommand{\av}[1]{\langle\, #1\, \rangle}

\newcommand{\half}{\frac{1}{2}}
\newcommand{\dif}{d}
\newcommand{\hc}{\mathrm{h.c.}}

\usepackage{scalerel,stackengine}
\newcommand\pig[1]{\scalerel*[5pt]{\big#1}{%
\ensurestackMath{\addstackgap[1.5pt]{\big#1}}}}
\newcommand\pigl[1]{\mathopen{\pig{#1}}}
\newcommand\pigr[1]{\mathclose{\pig{#1}}}

\newcommand{\cc}{c}
\newcommand{\nn}{n}
\newcommand{\mm}{m}
\newcommand{\GG}{G}
\newcommand{\FF}{F}
\newcommand{\MM}{M}
\newcommand{\KK}{K}
\newcommand{\LL}{L}

\newcommand{\tunnel}{J}
\newcommand{\NCG}{U}
\newcommand{\updownspin}{\mathord{\uparrow,\xssp\downarrow}}
\newcommand{\upspin}{\uparrow}
\newcommand{\downspin}{\downarrow}
\newcommand{\mym}{{\scalebox{0.85}{$\scriptstyle -$}}}
\newcommand{\myp}{{\scalebox{0.85}{$\scriptstyle +$}}}
\newcommand{\mypm}{{\scalebox{0.85}{$\scriptstyle \pm$}}}
\newcommand{\zzz}{z}

\newcommand{\CPHASE}{\mbox{\footnotesize \textrm{CPHASE}}}

\newcommand{\iSWAP}{\mbox{\footnotesize \textrm{iSWAP}}}

\newcommand{\FSIM}{\mbox{\small $\mathrm{FSIM}$}}

\definecolor{u0BgColor}{HTML}{ffedd4}
\definecolor{uXBgColor}{HTML}{deedfa}

\DeclareMathOperator{\diag}{diag}
\DeclareMathOperator{\sgn}{sgn}
\DeclareMathOperator{\tr}{tr}
\DeclareMathOperator{\arccot}{arccot}

\begin{document}

\title{Observation of separated dynamics of charge and spin in the Fermi-Hubbard model}

\author{Google AI Quantum and collaborators}

\email[Corresponding author (Z.~Jiang): ]{qzj@google.com}
\email[\\Corresponding author (V.~Smelyanskiy): ]{smelyan@google.com}

\date{\today}
\begin{abstract}
Strongly correlated quantum systems give rise to many exotic physical phenomena, including high-temperature superconductivity.  Simulating these systems on quantum computers may avoid the prohibitively high computational cost incurred in classical approaches.  However, systematic errors and decoherence effects presented in current quantum devices make it difficult to achieve this.  Here, we simulate the dynamics of the one-dimensional Fermi-Hubbard model using 16 qubits on a digital superconducting quantum processor.  We observe separations in the spreading velocities of charge and spin densities in the highly excited regime, a regime that is beyond the conventional quasiparticle picture.  To minimize systematic errors, we introduce an accurate gate calibration procedure that is fast enough to capture temporal drifts of the gate parameters.  We also employ a sequence of error-mitigation techniques to reduce decoherence effects and residual systematic errors.  These procedures allow us to simulate the time evolution of the model faithfully despite having over 600 two-qubit gates in our circuits.  Our experiment charts a path to practical quantum simulation of strongly correlated phenomena using available quantum devices.  
\end{abstract}

\maketitle

\begin{bibunit}[apsrev4-1_with_title]

The deceivingly simple Fermi-Hubbard model has greatly advanced our understanding of superconductivity, superfluidity, and quantum magnetism in correlated materials~\cite{lee_doping_2006,imada_metal-insulator_1998}.  The model is extremely hard to solve on classical computers in certain regimes, and it is widely used to benchmark numerical methods for strongly correlated systems~\cite{simons_collaboration_on_the_many-electron_problem_solutions_2015}.  A remarkable property of the one-dimensional (1D) Fermi-Hubbard model is spin-charge separation, i.e., spin and charge excitations travel at different speeds due to interparticle interactions~\cite{voit_one-dimensional_1995, jagla_numerical_1993, ulbricht_is_2009}.  Signatures of spin-charge separation were observed in solid state systems using angle-resolved photoemission spectroscopy~\cite{kim_observation_1996, segovia_observation_1999, kim_distinct_2006} and tunneling spectroscopy~\cite{tserkovnyak_interference_2003, auslaender_spin-charge_2005, jompol_probing_2009} as well as in cold-atom systems~\cite{ salomon_direct_2019,vijayan_time-resolved_2020} using site-resolved quantum gas microscopy~\cite{boll_spin-_2016, parsons_site-resolved_2016}.  

Here, we simulate the dynamics of an 8-site 1D Fermi-Hubbard model on a programmable superconducting quantum processor by suddenly removing the trapping potentials and turning on the on-site interactions.  We observe separations in the spreading velocities of spin and charge in a regime beyond the low-energy physics described by the Luttinger liquid theory~\cite{haldane_luttinger_1981,imambekov_one-dimensional_2012}.  Our platform enjoys the flexibility that analog devices do not, such as the ability to measure arbitrary observables, reverse the time evolution, and prepare for various initial states including BCS-type states~\cite{jiang_quantum_2018}.  It also has high repetition rates compared to some other platforms and avoids finite temperature effects. 

The key technical advancement that enabled this experiment is a calibration protocol that we recently developed for entangling gates, which we call Floquet calibration.  It allows for gate parameters to be characterized rapidly and precisely, which unlocks the ability to compensate for systematic errors caused by drifts and fluctuations.  Floquet calibration is based on the idea that an entangling gate can be uniquely determined by the eigenvalues of the composite gates consisting of the entangling gate and several different sets of single-qubit gates; one gets different snapshots of the entangling gate by changing the parameters in the single-qubit gates.  It is also robust to state preparation and measurement errors, sharing many common traits with robust phase estimation for calibrating single-qubit gates~\cite{kimmel_robust_2015, rudinger_experimental_2017}.  We also show that decoherence effects can be drastically suppressed by using a combination of error mitigation schemes.  Along with our calibration tool, these techniques allow us to increase the circuit depths (evolution times) by an order of magnitude.  Our work paves the way toward simulating strongly correlated quantum  systems~\cite{jordan_quantum_2012,georgescu_quantum_2014,barends_digital_2015,wecker_solving_2015,hensgens_quantum_2017,kivlichan_quantum_2018,jiang_quantum_2018,smith_simulating_2019,chiaro_growth_2019,mcardle_quantum_2020, rahmani_creating_2020, dallaire-demers_application_2020} on existing digital quantum computers.

\section{The model}
\label{sec:model}

\begin{figure*}[htb]
\centering
    \includegraphics[width=0.98\textwidth]{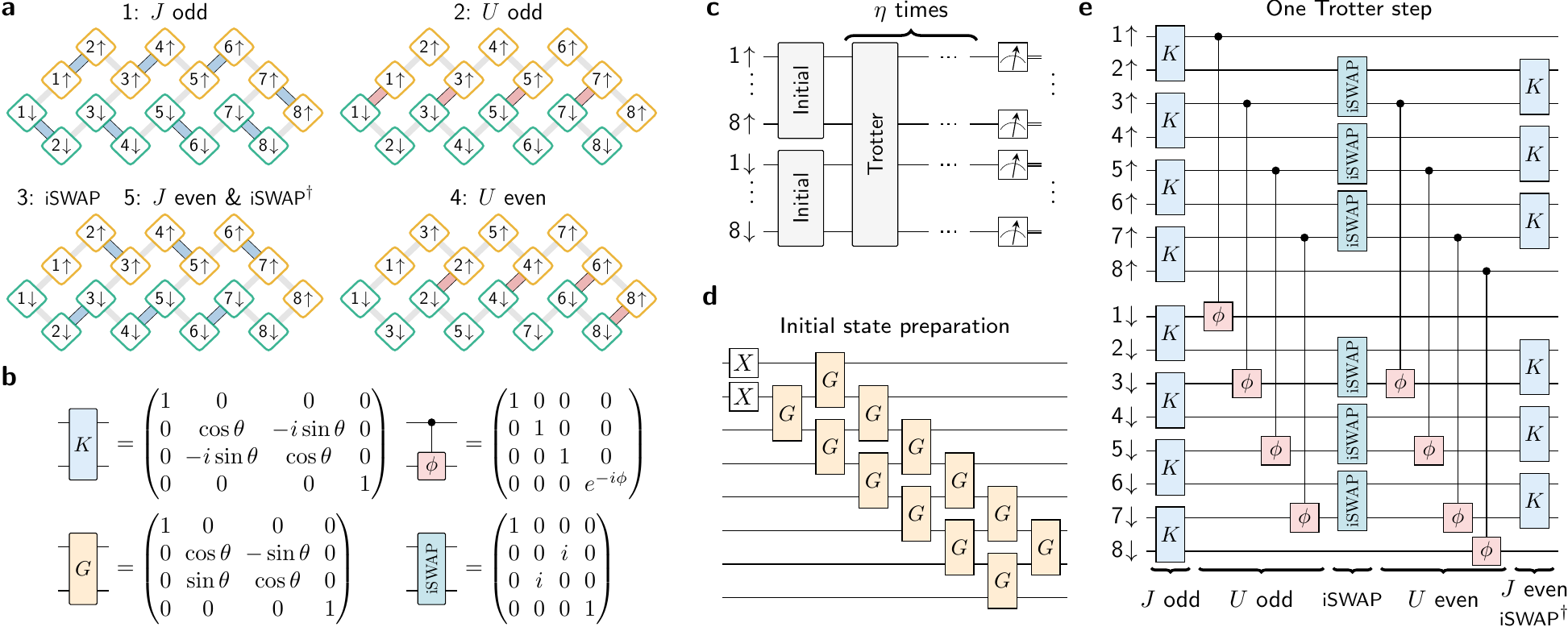}
    \caption{\textbf{Qubit layouts and quantum circuits.}  \textbf{a.} The 5 stages in a Trotter step, where the orange and green qubits represent spin-up and -down fermionic sites, respectively.  In stage 1 and 2, the blue and red edges represent the hopping and interaction terms, respectively.  In stage 3, we change the positions of the odd and even sites by applying the \iSWAP\ gates across the blue edges, which allows for implementing $U$ even in the next stage.  In stage 5, we swap the sites back to their original positions using the $\iSWAP^\dag$ gates, which are combined with the $J$ even terms. \textbf{b.} The matrix representations of the two-qubit gates. The Givens rotation gate $\GG$ (yellow) is used to prepare the initial state.  The \iSWAP-like gate $\KK$ (blue) and the \CPHASE\ gate (red) are used to implement the time evolution under the hopping and interaction terms, respectively.  The \iSWAP\ gate (green) is a special case of $\KK(\theta)$, with $\theta = \mathord{-}\pi/2$.  In Supplementary Fig.~\hyperref[{fig:sqrt_decompositions}]{\ref{fig:sqrt_decompositions}}, we show that any of the four gates can be decomposed into two $K(\pi/4)$ gates and several single-qubit gates.  \textbf{c.} The entire quantum circuit includes an initialization part, $\eta$ Trotter steps, and measurements in the Pauli-$Z$ basis. \textbf{d.} The circuit to prepare the ground state of a noninteracting Hamiltonian with two particles (excitations), where the angles of the Givens rotations can be determined using an efficient classical algorithm.  \textbf{e.} The quantum circuit to implement one Trotter step of the model.
  }
   \label{fig:circuits_layouts}
\end{figure*}

Consider the 1D Fermi-Hubbard model on $\LL$ lattice sites with open-boundary conditions,
\begin{align}\label{eq:h}
   H &= -\tunnel\sum_{j=1}^{\LL-1}\sum_{\nu=\updownspin }\!  \cc_{j,\nu}^\dagger\cc_{j+1,\nu} +\hc\nonumber\\ 
   &\quad + U\sum_{j=1}^{\LL} \nn_{j,\upspin}\xssp \nn_{j, \downspin} + \sum_{j=1}^{\LL}\sum_{\nu=\updownspin} \epsilon_{j, \nu}\, \nn_{j, \nu}\,,
\end{align}
where $\cc_{j,\nu}$ ($\cc_{j,\nu}^\dagger$) are the fermionic annihilation (creation) operators associated to site number $j$ and spin state $\nu$, and $\nn_{j,\nu}=\cc_{j,\nu}^\dagger\cc_{j,\nu}$ are the number operators.  The hopping term with coefficient $\tunnel$ in Eq.~\eqref{eq:h} describes particles tunneling between neighboring sites, the onsite interaction term with coefficient $U$ introduces an energy difference for doubly occupied sites, and the term $\epsilon_{j, \nu}$ represents spin-dependent local potentials.  The charge and spin densities are defined as the sum and difference of the spin-up and -down particle densities, respectively,
\begin{align}\label{eq:change_spin_density}
\rho_j^\pm = \av{\nn_{j, \upspin}} \pm \av{ \nn_{j, \downspin}}\,.
\end{align}
We map the fermionic operators to qubit operators using the Jordan-Wigner transformation (JWT) for each spin state, $\cc_{j,\nu} \;\mapsto\; \half\, (X_{j,\nu}+i Y_{j, \nu})\, Z_{1,\nu}\cdots Z_{j-1, \nu}$, where $X_{j, \nu}$, $Y_{j, \nu}$, and $Z_{j, \nu}$ are the Pauli operators.  Under the JWT, the unoccupied and occupied spin orbitals are represented by the qubit states $\ket{0}$ and $\ket{1}$, respectively.  We use the product formula~\cite{childs_nearly_2019}, i.e., Trotter steps, to simulate the time evolution of the system, where each term in the Hamiltonian~\eqref{eq:h} is implemented separately.  A single Trotter step is implemented with the 5 stages depicted in Fig.~\hyperref[{fig:circuits_layouts}]{\ref{fig:circuits_layouts}a}, where each spin state of the model is mapped to a zigzag chain of $8$ qubits; this optimizes the circuit depths under the geometric constraints.

\begin{figure*}[ht]\label{fig:scs}
   \includegraphics[width=0.99\textwidth]{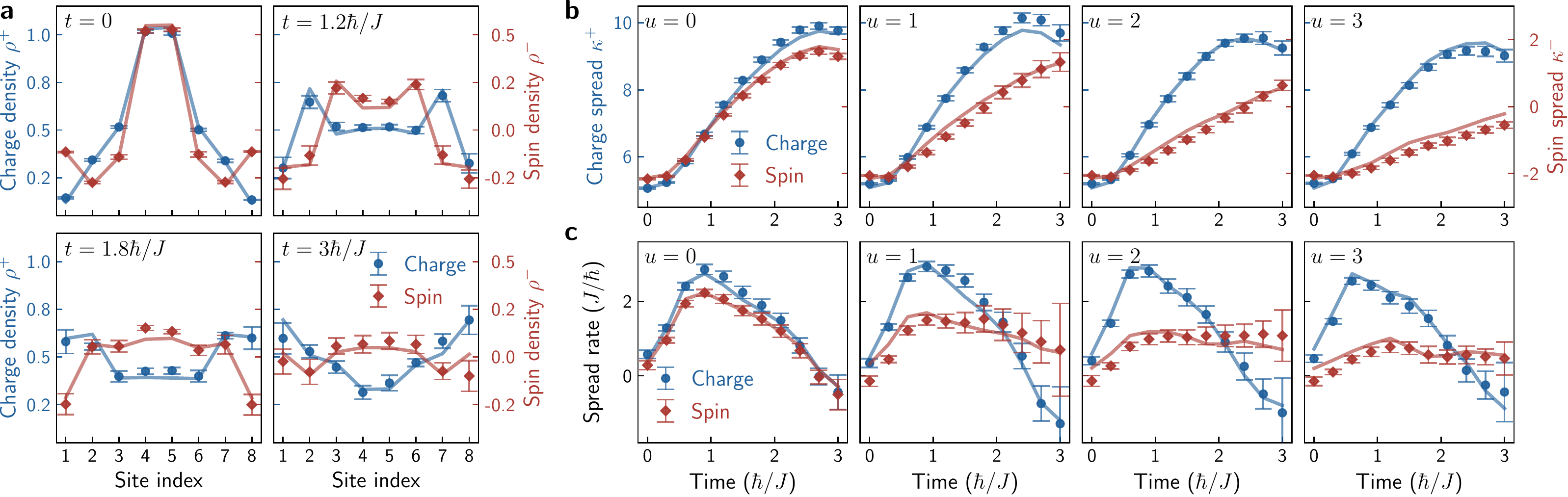}
   \caption{\textbf{Separation in charge and spin densities.}  We initialize the quantum state with $N_\uparrow = N_\downarrow=2$ (quarter filling), where the charge and spin densities $\rho_j^\pm = \av{\nn_{j, \upspin}} \pm \av{ \nn_{j, \downspin}}$ are peaked around the middle sites.  We then evolve the state under the Fermi-Hubbard Hamiltonian~\eqref{eq:h} with Trotter step length $\tau=0.3\xssp\hbar/\tunnel$.   Points and solid lines represent experimental and numerical (exact) results, respectively.  {\textbf a.} Time evolved charge (blue) and spin (red) densities for $u\equiv U/\tunnel=3$ with $t\xssp \tunnel/\hbar =0,1.2,1.8,3$ (corresponding to Trotter numbers $\eta = 0, 4, 6, 10$), where the error bars represent the standard error of the mean over 16 simulations with different choices of qubits and their arrangements, see Supplementary Fig.~\ref{fig:different_layouts}; the uncertainties due to finite sample sizes are much smaller (omitted on the plots).  The charge density spreads faster than the spin density and reaches the boundaries earlier.  {\textbf b.} The charge and spin spreads $\kappa^\pm=\sum_{j}\,\normb{ j-(\LL+1)/2} \, \rho_{j}^\pm$ as functions of the evolution time.  For $u=0$, they almost lay on top of each other; the small discrepancy is due to the parasitic \CPHASE\ in our native gate.  In comparison, they are well separated for larger interaction strengths $u\geq 1$.
  {\textbf c.} The numerical derivatives of $\kappa^\pm$ with respect to the evolution time.
 }
\end{figure*}

Under the JWT, the hopping term $\cc_{j,\nu}^\dagger\cc_{j+1,\nu} +\hc$ is mapped to $\frac{1}{2}\xssp (X_{j,\nu} X_{j+1,\nu} + Y_{j,\nu} Y_{j+1,\nu})$.  Its time evolution can be implemented using the two-qubit gate $K(\theta)$ in Fig.~\hyperref[{fig:circuits_layouts}]{\ref{fig:circuits_layouts}b}.  This gate is used in the first and last stages in the circuit depicted in Fig.~\hyperref[{fig:circuits_layouts}]{\ref{fig:circuits_layouts}e}.  In the first stage, we set $\theta = -\tau\xssp\tunnel/ \hbar$ to implement a time step of length $\tau$.  In the last stage, we set $\theta = -\tau J/\hbar+\pi/2$, where the extra angle $\pi/2$ is used to undo the \iSWAP\ gates in the third stage to change the positions of the fermionic sites.  The gate $K(\theta)$ with arbitrary $\theta$ can be decomposed into two $\KK(\pi/4)=\sqrt{\iSWAP}^{\,\dag}$ gates and several single-qubit $Z$ rotations, see Supplementary Fig.~\hyperref[{fig:sqrt_decompositions}]{\ref{fig:sqrt_decompositions}c}.  Our hardware native two-qubit gate takes the form $\KK(\vartheta)\,\CPHASE\xssp(\varphi)$, where $\vartheta\approx \pi/4$ and the parasitic controlled phase $\varphi\lesssim \pi/20$.  At the time we took the data, the means and standard deviations of the two parameters across different pairs of qubits were \SI[parse-numbers = false]{\vartheta= 0.783 \pm 0.012}{\radian} and \SI[parse-numbers = false]{\varphi = 0.138 \pm 0.015}{\radian}.  The $\CPHASE\xssp(\varphi)$ term introduces an interaction term between neighboring fermionic sites $V \nn_{j, \nu}\,\nn_{j+1, \nu}$ with $V = 2 \hbar \varphi /\tau$.  It has sizable effects for longer evolution times, and we include it in our numerical simulations to compare to experimental results.  The entanglement part in our native two-qubit gate takes about \SI{12}{\nano\second} and is preceded by single-qubit $Z$ rotations, which take \SI{10}{\nano\second} with a \SI{5}{\nano\second} padding on each side.  Therefore, one hopping term takes about $2\times \SI{32}{\nano\second} = \SI{64}{\nano\second}$ to implement on the hardware. 
  
The time evolution of the on-site interaction term $\nn_{j,\upspin}\xssp \nn_{j, \downspin}$
can be implemented using $\CPHASE(\phi)$ gate with $\phi = \tau U/\hbar $.  It can be decomposed exactly into two native two-qubit gates and single-qubit $X$ and $Z$ rotations, see Supplementary Information~\ref{sec:gate_decompositions}.  There are three layers of $X$ rotations (microwave gates) in the composite gate, each taking about \SI{25}{\nano\second}.  Therefore, the entire composite \CPHASE\ gate takes about \SI{139}{\nano\second} to implement.  As shown in Fig.~\hyperref[{fig:circuits_layouts}]{\ref{fig:circuits_layouts}a}, we implement the \CPHASE\ gate on the odd and even sites separately due to geometry constraints.  Idling qubits are susceptible to crosstalk and low-frequency noises in the $Z$ basis, and we mitigate them by applying spin echos consisting of pairs of $X$ gates (not shown in Fig.~\ref{fig:circuits_layouts}). 

We initialize the system into the ground state of a non-interacting fermionic Hamiltonian using networks of Givens rotations~\cite{wecker_solving_2015}, i.e., two-mode fermionic basis transformations.  The Givens rotation takes the matrix form $G$ in Fig.~\hyperref[{fig:circuits_layouts}]{\ref{fig:circuits_layouts}b} when acting on neighboring qubits.  It can also be decomposed into two $\KK(\pi/4)$ gates and single-qubit $Z$ rotations, see Supplementary Fig.~\hyperref[{fig:sqrt_decompositions}]{\ref{fig:sqrt_decompositions}b}.  By parallelizing the Givens rotations, the ground state of an arbitrary $\LL$-mode non-interacting Hamiltonian can be prepared in circuit depth $\mathcal O(\LL)$ ~\cite{kivlichan_quantum_2018, jiang_quantum_2018}.  Recently, the Givens rotation network was successfully used to variationally construct a chemically-accurate  Hartree-Fock state~\cite{collaborators_hartree-fock_2020}. Here we use the OpenFermion code~\cite{mcclean_openfermion_2020} based on the scheme in~\cite{jiang_quantum_2018},  which requires $\mathord\sim \LL^2/4$ Givens rotations with circuit depth $\mathord\sim \LL$ near half filling.  In Fig.~\hyperref[{fig:circuits_layouts}]{\ref{fig:circuits_layouts}d}, we plot the initialization circuit for two fermions. 

\section{Separation of spin and charge velocities}
\label{sec:separation}

In the Luttinger liquid description of the 1D Fermi-Hubbard model~\cite{luttinger_exactly_1963, mattis_exact_1965, lieb_absence_1968}, low-energy charge and spin excitations propagate at different characteristic velocities; see Supplementary Information~\ref{sec:spin_charge_boson}.  It is based on the assumption that the system is close to its ground state.  Here we observe separations in the dynamics of charge and spin densities in a highly excited regime, where the Luttinger liquid theory does not formally apply.

Consider an 8-site 1D Fermi-Hubbard system with $N_{\nu}$ particles in the spin state $\nu$.  We prepare the initial state $\ket{\psi_0}$ as the ground state of an non-interacting Hamiltonian $H_0$ by setting $U=0$ in Eq.~\eqref{eq:h}.  The local potentials in $H_0$ are chosen to have a Gaussian form $\epsilon_{j,\nu} =  - \lambda_\nu\, e^{- \half \xssp (j - m_\nu)^2 /  \sigma_\nu^2}$,
where $\lambda_\nu$, $m_\nu$, and $\sigma_\nu$ set the magnitude, center, and width of the potentials, respectively.  We set the parameters of the spin-up Gaussian potential to $\lambda_\upspin = 4$, $m_\upspin = 4.5$, and $\sigma_\upspin = 1$ while leaving the spin-down potential to zero.  This generates initial charge and spin density peaks in the middle of the chain, see subplot $t=0$ in Fig.~\hyperref[{fig:scs}]{\ref{fig:scs}a}. 

\begin{figure*}[ht]\label{fig:two_gaussians}
   \includegraphics[width=0.99\textwidth]{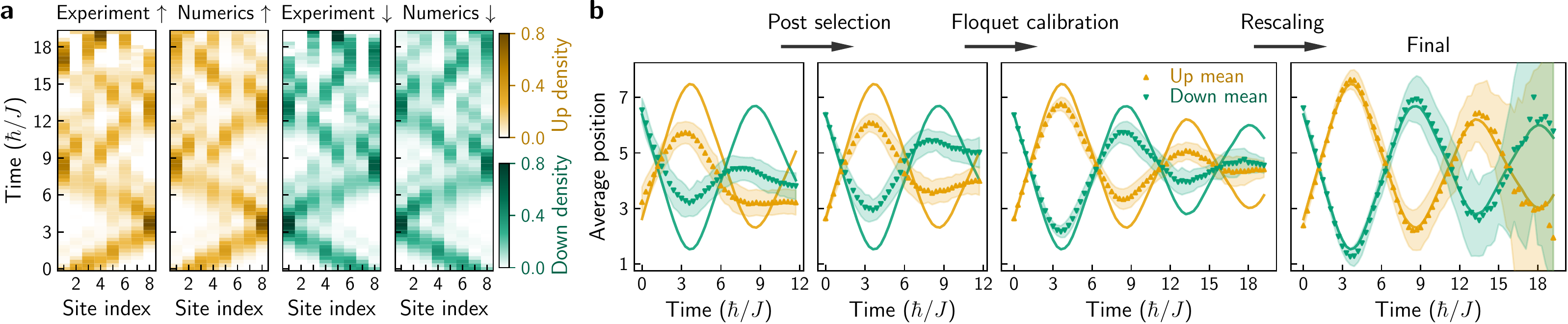}
   \caption{\textbf{Noninteracting time evolution and error mitigation.} 
  \textbf{a.} Time evolution of the particle densities $\av{\nn_{j,\xssp \nu}}$, where $j=1,\ldots, 8$ and $\nu = \uparrow, \downarrow$.  The experiment results match well with the numerics (exact) for $t\lesssim 16.5\xssp \hbar/J$, corresponding to a circuit depth of 228 layers of two-qubit gates with execution time \SI{7.3}{\mu\second}.  \textbf{b.} Demonstrating error mitigation schemes using the average positions $\sum_j \av{\nn_{j,\xssp \nu}}\, j$, where the solid lines represent numerical results for the spin-up (yellow) and -down (green) particles.  The triangles and the shaded areas represent the means and sample standard deviations over 16 simulations with different choices of qubits and their arrangements, see Supplementary Fig.~\ref{fig:different_layouts}.  This procedure, which we call qubit assignment averaging, removes inhomogeneous effects in the system.  Decoherence due to loss of excitations ($T_1$ errors) can be removed by postselecting the measurement results with the correct numbers of excitations.  Floquet calibration improves both the means and standard deviations of the average positions.   Crucially, the particle density distributions obtained after the calibration are similar to the exact solutions with damped amplitudes.  The damping factor is a function of the evolution time $t$; its value is approximately $1$ at $t=0$ and $0.16$ at $t = 16.5\xssp \hbar/J$.  This allows us to get excellent agreement with theory predictions by rescaling.
    }
\end{figure*}

We then evolve the system under the Hamiltonian~\eqref{eq:h} with the Trotter step described in Fig.~\hyperref[{fig:circuits_layouts}]{\ref{fig:circuits_layouts}e} by setting the time step length to $\tau = 0.3\xssp\hbar/\tunnel$ and the local potentials to $\epsilon_{j,\nu}=0$. In Fig.~\hyperref[{fig:scs}]{\ref{fig:scs}a}, we plot the distributions of the charge and spin densities at several evolution times for $N_\uparrow=N_\downarrow=2$ and $u\equiv U/\tunnel=3$, where the dynamics of the charge and spin degrees of freedom are separated.  We leave more detailed results for this case and the $N_\uparrow = N_\downarrow=3$ case in Supplementary Figs.~\ref{fig:full_data_2_2} and \ref{fig:full_data_3_3}, respectively.

To quantify the degree that charge and spin densities spread from the middle of the chain, we introduce
\begin{align} \label{eq:spread}
  \kappa_\eta^\pm =  \sum_{j=1}^{\LL}\,\pigl\lvert j-(\LL + 1)/2\pigr\rvert \, \rho_{j,\eta}^\pm \,,
\end{align}
where $\rho_{j,\eta}^\pm = \av{\nn_{j, \upspin}} \pm \av{ \nn_{j, \downspin}}$ are the charge and spin densities after $\eta$ Trotter steps.  In Fig.~\hyperref[{fig:scs}]{\ref{fig:scs}b}, we plot $\kappa^\pm$ as functions of the evolution time for several interaction strengths $u$.  When $u=0$, they nearly coincide with each other; the small separation is the result of the nearest-neighbor interaction term $V \nn_{j, \nu}\,\nn_{j+1, \nu}$, caused by the parasitic \CPHASE\ in our native gate.  The gaps between $\kappa^+$ and $\kappa^-$ widen as $u$ increases, demonstrating increased separations as the system goes into the strongly interacting regime.  The charge spread $\kappa^+$ reaches the maximum value after it has fully hit the boundaries.  In comparison, $\kappa^-$ increases at significantly lower rates for all $u\neq 0$ and never fully reaches the boundaries within the maximum evolution time.

In Fig.~\hyperref[{fig:scs}]{\ref{fig:scs}c}, we plot the numerical derivatives of the charge and spin spreads $\kappa^\pm$.  Because our initial wavefunction is real, both charge and spin currents equal zero at $t=0$.  The small nonzero values of the observed initial rates of charge and spin spreads are due to the finite Trotter step length and the parasitic \CPHASE\ in the initialization circuit.  The charge spreading rate gradually increases until the particles starts to hit the boundaries.  As the interaction strength $u$ increases, the maximum spin spread rate decreases.  In comparison, the maximum charge spread rate roughly keeps the same.

\section{Error mitigation and calibration}
\label{sec:error_mitigation}

To reach the desired circuit depths, we employ a combination of error mitigation and calibration schemes.  To illustrate that, we consider the case where there is exactly one particle in each spin state $N_\uparrow = N_\downarrow=1$.  We initialize the spin-up (down) particle into a left (right)-moving Gaussian wavepacket.  This can be achieved by first creating a real Gaussian wavepacket with the Givens rotations and then generating a phase gradient using single-qubit $Z$ rotations.  We evolve the system under the hopping terms by including only the first and last stages in Fig.~\hyperref[{fig:circuits_layouts}]{\ref{fig:circuits_layouts}e} with time step length $\tau = 0.3\hbar/\tunnel$.  In Fig.~\hyperref[{fig:two_gaussians}]{\ref{fig:two_gaussians}a}, we compare the numerics with experimental results after applying all the error mitigation schemes.  They match well with each other up to $t\approx 16.5\xssp\hbar/\tunnel$ (or $55$ Trotter steps), and the clear interference patterns at larger times indicate that the evolution is coherent instead of diffusive.  The two spin states evolve independently here, i.e., there is no gate between the corresponding sets of qubits.  However, we can account for the effects of crosstalks more properly by including the two spin states.  The impact of each step in our error mitigation procedure is illustrated in Fig.~\hyperref[{fig:two_gaussians}]{\ref{fig:two_gaussians}b}, which we describe below one by one.

\begin{figure}[t]\label{fig:pss}
\centering
    \includegraphics[width=0.48\textwidth]{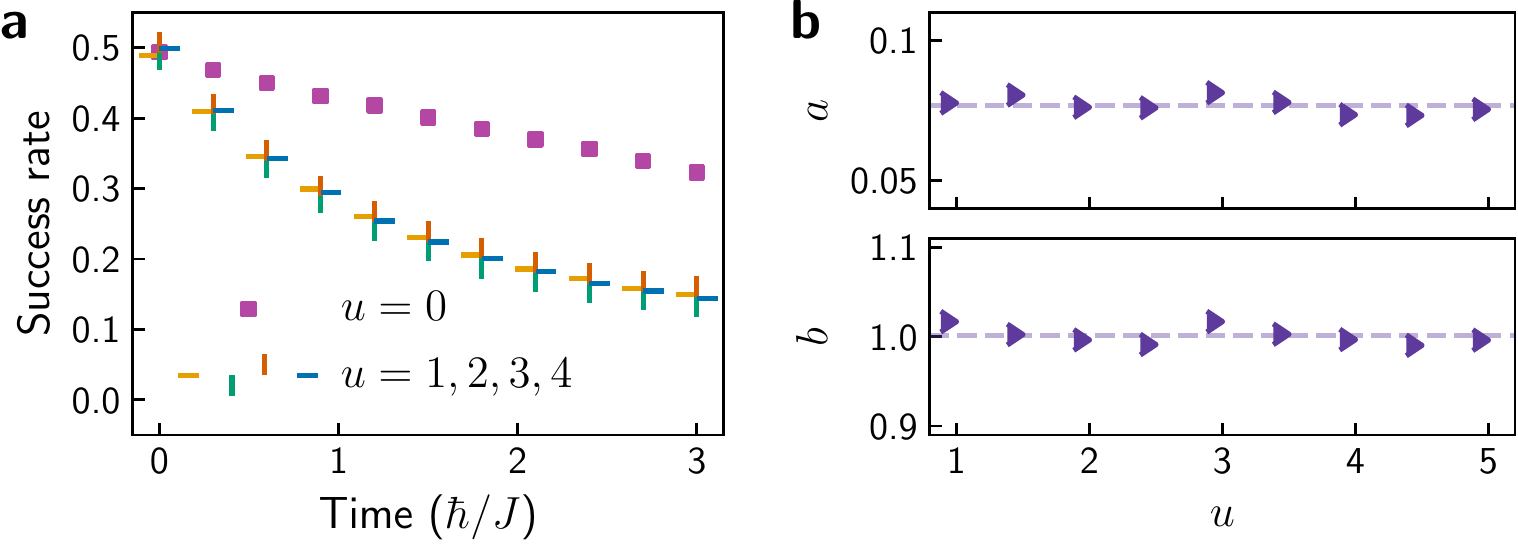}
    \caption{\textbf{Postselection success rate and rescaling parameters.}
    {\textbf a.}  The post selection success rates as functions of the evolution time.  The squares denote the results for the noninteracting case in Fig.~\ref{fig:two_gaussians}, which decays much slower due to the reduced number of gates in each Trotter step.  The horizontal and vertical bars denote results for the interacting case in Fig.~\ref{fig:scs} with different values of $u$; the cross formation of the horizontal and vertical bars shows that the success rate does not depend on $u$.  {\textbf b.}  The rescaling parameters $a$ and $b$ in Eq.~\eqref{eq:rescale} for the case in Fig.~\ref{fig:scs}.  Their values barely depend on the interaction strength $u$, allowing one to infer them by comparing numerical and experimental results in the weakly interacting regime.}
    \label{fig:post_selection_rate}
\end{figure}

The quantum circuit for the Fermi-Hubbard model conserves the total number of excitations for each spin state.  However, excitations can leave the system due to interactions with the environment.  In Supplementary Fig.~\ref{fig:t1}, we plot the $T_1$ map of the device at the time that the final data were taken.  This kind of error can be removed by postselecting the measurement results with the correct numbers of excitations.  It also reduces state preparation and measurement (SPAM) errors to some extent.  The postselection success rates of the noninteracting instance shown in Fig.~\ref{fig:two_gaussians} are about $0.5$ for $\eta = 0$ and $0.2$ for $\eta = 55$, where $\eta$ is the number of Trotter steps.  In Fig.~\hyperref[{fig:pss}]{\ref{fig:pss}a}, we plot the success rates as functions of the evolution time.  We take 20,000 samples for each circuit to make sure that the uncertainties of the expectation values are low after postselection.  

Systematic errors are especially detrimental to quantum computation as their effects can add up coherently.  In addition to our routine calibration, we also use Floquet calibration, a fast calibration method that we recently developed for characterization of entangling gates.  To capture crosstalks, we calibrate the two-qubit gates in each configuration in Fig.~\hyperref[{fig:circuits_layouts}]{\ref{fig:circuits_layouts}a} by applying them simultaneously; the results are then used to correct the two-qubit gates in the quantum circuit within the same configuration.  Floquet calibration can determine most parameters of the two-qubit gates to an uncertainty of less than \SI[product-units = single]{e-3}{\radian} under one minute, sufficiently fast to characterize errors due to drifts and fluctuations in the control fields and qubit frequencies.  It is also robust to state preparation and measurement (SPAM) errors in general.  In Supplementary Fig.~\ref{fig:floquet_angles}, we plot the parameters in our native two-qubit gate obtained using Floquet calibration during a period of several hours.  We also plot the two-qubit gate fidelity map from our routine calibration using cross-entropy benchmarking (XEB) in Supplementary Fig.~\ref{fig:xeb}; more about XEB and our calibration process can be found in Supplementary Information in Ref.~\cite{arute_quantum_2019}.  We leave detailed discussions on the implementations and properties of Floquet calibration in Supplementary Information~\ref{sec:floq_calib}.

Inhomogeneities in gate parameters and decoherence rates are common in quantum computing devices.  They make the experiment results unpredictable and implementation dependent.  We solve this issue using qubit assignment averaging, where experiment results are averaged over 16 different realizations, see Supplementary Fig.~\ref{fig:different_layouts}.  In each realization, we either choose a different set of qubits, arrange the qubits differently, or do both.  In Fig.~\hyperref[{fig:two_gaussians}]{\ref{fig:two_gaussians}b}, we show that the averaged results are smooth even if the outcomes from individual implementations fluctuate significantly.  More importantly, averaging makes it possible to describe the simulation results using models with randomized parameters.  The values of the observables are often damped in a predictable way in these models, making it possible for further mitigation of the errors.

Finally, we rescale the damped expectation values $\av{\nn_{j, \nu}}$, leading to excellent agreement with theoretical predictions.  We choose the fiducial point for rescaling to be $\bar\nn_{\nu} = N_\nu / \LL$, i.e., the averaged particle density for the spin state $\nu$.  We observe that the damping factor is approximately linear in the number of Trotter steps $\eta$,
\begin{align}
\frac{\av{\nn_{j, \nu}}_\mathrm{exp}- \bar\nn_{\nu}}{\av{\nn_{j, \nu}}_\mathrm{num}- \bar\nn_{\nu}} \approx b -a \eta\,,
\label{eq:rescale}
\end{align}
where exp and num stand for experimental and numerical results, respectively.  The parameter $a$ ($b$) describes the damping effect of the Trotter steps (initial state preparation circuit).  The linear relation fits the experimental results well when the damping factor is $\gtrsim 0.2$ for both noninteracting and interacting cases, see Supplementary Fig.~\ref{fig:scaling}.  The fitted values of $a$ and $b$ hardly depend on the interaction strength $U$, see Fig.~\hyperref[{fig:pss}]{\ref{fig:pss}b}.  Therefore, we can estimate their values by comparing the experimental and numeric results in a regime that is easy to solve classically, e.g., the weak interaction regime.  The linear relation~\eqref{eq:rescale} is not essential to our rescaling procedure.  However, the weak dependence of the damping factor on the interaction strength $U$ is crucial.  

\begin{table}[t]
    \centering
    \newcommand{\npad}{0.3em}
    \renewcommand{\tabcolsep}{0pt}
    \renewcommand{\arraystretch}{1.2}
    
\begin{tabular}{c|@{}c@{}}\hline
Case
&
 \begin{tabular}{r r r r r r}
        \multicolumn{1}{>{\centering\arraybackslash}p{11.38mm}}{$t_\mathrm{evol}$ ($\hbar/\nsp\tunnel$)}&
        \multicolumn{1}{>{\centering\arraybackslash}p{11.49mm}}{$t_\mathrm{circuit}$ (\si{\mu\second})} &
        \multicolumn{1}{>{\centering\arraybackslash}p{12.21mm}}{Circuit depths} &
        \multicolumn{1}{>{\centering\arraybackslash}p{12.20mm}}{2-qubit counts} &
        \multicolumn{1}{>{\centering\arraybackslash}p{11.69mm}}{$\mu$-wave counts} &
        \multicolumn{1}{>{\centering\arraybackslash}p{11.8mm}}{$R_Z$ counts}
\end{tabular}
\tabularnewline\hline\hline
\rowcolor{uXBgColor!70!white}
\parbox{1.50cm}{$U\neq0$ \\ $N_{\nsp P} \nsp=\nsp 4,\nsp 6$ }
&
\begin{tabular}{r r r r r r}
\rowcolor{uXBgColor!70!white}
        \multicolumn{1}{>{\centering\arraybackslash}p{11.38mm}}{1.5}&
        \multicolumn{1}{>{\centering\arraybackslash}p{11.49mm}}{2.8}&
        \multicolumn{1}{>{\centering\arraybackslash}p{12.21mm}}{159} &
        \multicolumn{1}{>{\centering\arraybackslash}p{12.20mm}}{328} &
        \multicolumn{1}{>{\centering\arraybackslash}p{11.69mm}}{364}&
        \multicolumn{1}{>{\centering\arraybackslash}p{11.8mm}}{566}
        \\
        \hline
        \rowcolor{uXBgColor!70!white}
         \multicolumn{1}{>{\centering\arraybackslash}p{11.38mm}}{3.0} 
        &\multicolumn{1}{>{\centering\arraybackslash}p{11.49mm}}{5.2}&
        \multicolumn{1}{>{\centering\arraybackslash}p{12.21mm}}{289}&
        \multicolumn{1}{>{\centering\arraybackslash}p{12.20mm}}{608} &
        \multicolumn{1}{>{\centering\arraybackslash}p{11.69mm}}{724}&
        \multicolumn{1}{>{\centering\arraybackslash}p{11.8mm}}{1056}
\end{tabular}
\tabularnewline\hline
\rowcolor{u0BgColor!70!white}
\parbox{1.50cm}{$U = 0$ \\ $N_{\nsp P} \nsp=\nsp 2$ }
&
\begin{tabular}{r r r r r r}
\rowcolor{u0BgColor!70!white}
        \multicolumn{1}{>{\centering\arraybackslash}p{11.38mm}}{9.0} 
        &\multicolumn{1}{>{\centering\arraybackslash}p{11.49mm}}{4.1}&
        \multicolumn{1}{>{\centering\arraybackslash}p{12.21mm}}{257}&
        \multicolumn{1}{>{\centering\arraybackslash}p{12.20mm}}{\:$434\scalebox{1}{$\times 2$}$} &
        \multicolumn{1}{>{\centering\arraybackslash}p{11.69mm}}{2}&
        \multicolumn{1}{>{\centering\arraybackslash}p{11.8mm}}{\:$836\scalebox{1}{$\times 2$}$}
        \\
        \hline
        \rowcolor{u0BgColor!70!white}
        \multicolumn{1}{>{\centering\arraybackslash}p{11.38mm}}{16.5}
        &\multicolumn{1}{>{\centering\arraybackslash}p{11.49mm}}{7.3}&
        \multicolumn{1}{>{\centering\arraybackslash}p{12.21mm}}{457}&
        \multicolumn{1}{>{\centering\arraybackslash}p{12.20mm}}{\:$784\scalebox{1}{$\times 2$}$} &
        \multicolumn{1}{>{\centering\arraybackslash}p{11.69mm}}{2}&
        \multicolumn{1}{>{\centering\arraybackslash}p{11.8mm}}{\:$1511\scalebox{1}{$\times 2$}$}
\end{tabular}
\tabularnewline\hline
\end{tabular}
    \caption{\textbf{Circuit statistics.} Circuit statistics for the interacting case $U\neq 0$ and the noninteracting case $U=0$ with different numbers of particles $N_P$, where $t_\mathrm{evol}$ and $t_\mathrm{circuit}$ are the Hamiltonian evolution time and circuit execution time, respectively.  The circuit depths include the contributions from the two-qubit gates, microwave gates ($\mu$-wave), and single-qubit $Z$ rotations.  We also count the total numbers of the constituent gates.  The microwave gates are single-qubit rotations along axes on the $X$-$Y$ plane, which are used in the interaction terms and the initial state preparations.  The single-qubit $Z$ rotations ($R_Z$) are always bundled with our two-qubit gates and do not require extra time to implement. 
    }
    \label{tab:circuit_statistics}
\end{table}

\section{Conclusion and outlook}
\label{sec:conclusion}

Using a combination of the error mitigation and calibration schemes, we have extended our quantum circuits to unprecedented depths, see statistics in Table~\ref{tab:circuit_statistics}.  This opens the possibility of simulating strongly correlated systems on current quantum computing devices, such as the classically hard 2D Fermi-Hubbard model.  The recipe for error mitigation that we use here can also be useful to many other applications, including the variational quantum eigensolver (VQE)~\cite{mcclean_theory_2016} and quantum approximate optimization algorithm (QAOA)~\cite{farhi_quantum_2014, arute_quantum_2020}.  We also expect our calibration technique to play a central role in quantum device characterization and Hamiltonian learning.

\section{author contributions} 
Z. Jiang and V. Smelyanskiy designed the experiment;
Z. Jiang and W. Mruczkiewicz developed the code, collected the data, and wrote the paper;
L. Ioffe, K. Kechedzhi, and V. Smelyanskiy assisted with the physical theory of the model;
R. Babbush, S. Boixo, J. McClean, and N. Rubin contributed to the algorithmic part of the project;
Z. Jiang and V. Smelyanskiy developed the theory for Floquet calibration;
Y. Chen, Z. Jiang, W. Mruczkiewicz, C. Neill, M. Niu, and Xiao Mi implemented the Floquet calibration; S.~J.~Cotton,
C. Mejuto-Zaera, P. Schmitteckert, N. Tubman, and N. Vogt helped with numerical simulations of the model;
J. Gross, J. Martinis, P. Roushan, and X. Mi helped for improving the presentations of the results.  Experiments were performed---through Google's Quantum Computing Service---using a quantum processor that was recently developed and fabricated by a large effort involving the entire Google AI Quantum team.
 
\section{acknowledgements} 
Dave Bacon is a CIFAR Associate Fellow in the Quantum Information Science Program.  ZJ would like to thank Philipp Hauke for his helpful comments on the manuscript.  \textbf{Funding:} This work was supported by Google LLC.   N.M.T and S.J.C are grateful for support from NASA Ames Research Center as well as support from the AFRL Information Directorate under grant F4HBKC4162G001.  Some calculations were performed as part of the XSEDE computational Project No. TG-MCA93S030.  \textbf{Competing Interests:} The authors declare no competing interests.  Supplementary Information is available for this paper.  \textbf{Data and materials availability:} The code used for this experiment and a tutorial for running it can be found in the open source library ReCirq, located at \url{https://doi.org/10.5281/zenodo.4091471}. All data needed to evaluate the conclusions in the paper are present in the paper or the Supplementary Information. Data presented in the figures can be found in the Dryad repository located at \url{https://doi.org/10.5061/dryad.crjdfn32v}.

\onecolumngrid

\vspace{1em}
\begin{flushleft}
{\Large Google AI Quantum and Collaborators}

\bigskip

\renewcommand{\author}[2]{#1$\!^\textrm{\scriptsize #2}$}
\renewcommand{\affiliation}[2]{$^\textrm{\scriptsize #1}$ #2 \\}

\newcommand{\Google}{1}
\newcommand{\UMass}{2}
\newcommand{\Caltech}{3}
\newcommand{\UCSB}{4}
\newcommand{\NASA}{5}
\newcommand{\KBR}{6}
\newcommand{\Bosch}{7}
\newcommand{\Sherbrooke}{8}
\newcommand{\UCR}{9}
\newcommand{\HQS}{10}
\newcommand{\Oxford}{11}
\newcommand{\UCB}{12}
\newcommand{\UM}{13}

\newcommand{\xGoogle}{\affiliation{\Google}{Google Research}}
\newcommand{\xUMass}{\affiliation{\UMass}{Department of Electrical and Computer Engineering, University of Massachusetts, Amherst, MA}}
\newcommand{\xCaltech}{\affiliation{\Caltech}{Department of Physics, California Institute of Technology, Pasadena, CA}}
\newcommand{\xUCSB}{\affiliation{\UCSB}{Department of Physics, University of California, Santa Barbara, CA}}
\newcommand{\xSherbrooke}{\affiliation{\Sherbrooke}{Institut quantique and D\'epartment de Physique, Universit\'e de Sherbrooke, Qu\'ebec J1K 2R1, Canada}}
\newcommand{\xUCR}{\affiliation{\UCR}{Department of Electrical and Computer Engineering, University of California, Riverside, CA}}
\newcommand{\xHQS}{\affiliation{\HQS}{HQS Quantum Simulations GmbH, Haid-und-Neu-Stra\ss e 7,
76131 Karlsruhe, Germany}}
\newcommand{\xBosch}{\affiliation{\Bosch}{Robert Bosch GmbH, Robert-Bosch-Campus 1, 71272 Renningen, Germany}}
\newcommand{\xNASA}{\affiliation{\NASA}{Quantum Artificial Intelligence Laboratory (QuAIL), NASA Ames Research Center, Moffett Field, CA 
}}
\newcommand{\xKBR}{\affiliation{\KBR}{KBR, 601 Jefferson St., Houston, TX 77002}}
\newcommand{\xUCB}{\affiliation{\UCB}{Department of Chemistry, University of California, Berkeley, CA}}
\newcommand{\xUM}{\affiliation{\UM}{Department of Electrical Engineering and Computer Science, University of Michigan, Ann Arbor, MI}}
\newcommand{\xOxford}{\affiliation{\Oxford}{Department of Materials, University of Oxford, Parks Road, Oxford OX1 3PH, United Kingdom}}

\author{Frank Arute}{
\Google
}%
,
\author{Kunal Arya}{
\Google
}%
,
\author{Ryan Babbush}{
\Google
}%
,
\author{Dave Bacon}{
\Google
}%
,
\author{Joseph C.~Bardin}{
\Google,\!
\UMass
}%
,
\author{Rami Barends}{
\Google
}%
,
\author{Andreas Bengtsson}{
\Google
}%
,
\author{Sergio Boixo}{
\Google
}%
,
\author{Michael Broughton}{
\Google
}%
,
\author{Bob B.~Buckley}{
\Google
}%
,
\author{David A.~Buell}{
\Google
}%
,
\author{Brian Burkett}{
\Google
}%
,
\author{Nicholas Bushnell}{
\Google
}%
,
\author{Yu Chen}{
\Google
}%
,
\author{Zijun Chen}{
\Google
}%
,
\author{Yu-An Chen}{
\Google,\!
\Caltech
}%
,
\author{Ben Chiaro}{
\Google,\!
\UCSB
}%
,
\author{Roberto Collins}{
\Google
}%
,
\author{Stephen J. Cotton}{
\NASA,\!
\KBR
}%
,
\author{William Courtney}{
\Google
}%
,
\author{Sean Demura}{
\Google
}%
,
\author{Alan Derk}{
\Google
}%
,
\author{Andrew Dunsworth}{
\Google
}%
,
\author{Daniel Eppens}{
\Google
}%
,
\author{Thomas Eckl}{
\Bosch
}%
,
\author{Catherine Erickson}{
\Bosch
}%
,
\author{Edward Farhi}{
\Google
}%
,
\author{Austin Fowler}{
\Google
}%
,
\author{Brooks Foxen}{
\Google
}%
,
\author{Craig Gidney}{
\Google
}%
,
\author{Marissa Giustina}{
\Google
}%
,
\author{Rob Graff}{
\Google
}%
,
\author{Jonathan A.~Gross}{
\Google,\!
\Sherbrooke
}%
,
\author{Steve Habegger}{
\Google
}%
,
\author{Matthew P.~Harrigan}{
\Google
}%
,
\author{Alan Ho}{
\Google
}%
,
\author{Sabrina Hong}{
\Google
}%
,
\author{Trent Huang}{
\Google
}%
,
\author{William Huggins}{
\Google
}%
,
\author{Lev B.~Ioffe}{
\Google
}%
,
\author{Sergei V.~Isakov}{
\Google
}%
,
\author{Evan Jeffrey}{
\Google
}%
,
\author{Zhang Jiang}{
\Google
}%
,
\author{Cody Jones}{
\Google
}%
,
\author{Dvir Kafri}{
\Google
}%
,
\author{Kostyantyn Kechedzhi}{
\Google
}%
,
\author{Julian Kelly}{
\Google
}%
,
\author{Seon Kim}{
\Google
}%
,
\author{Paul V.~Klimov}{
\Google
}%
,
\author{Alexander N.~Korotkov}{
\Google,\!
\UCR
}%
,
\author{Fedor Kostritsa}{
\Google
}%
,
\author{David Landhuis}{
\Google
}%
,
\author{Pavel Laptev}{
\Google
}%
,
\author{Mike Lindmark}{
\Google
}%
,
\author{Erik Lucero}{
\Google
}%
,
\author{Michael Marthaler}{
\HQS}%
,
\author{Orion Martin}{
\Google
}%
,
\author{John M.~Martinis}{
\Google,\!
\UCSB
}%
,
\author{Anika Marusczyk}{
\Bosch
}%
,
\author{Sam McArdle}{
\Google,\!
\Oxford
}%
,
\author{Jarrod R.~McClean}{
\Google
}%
,
\author{Trevor McCourt}{
\Google
}%
,
\author{Matt McEwen}{
\Google,\!
\UCSB
}%
,
\author{Anthony Megrant}{
\Google
}%
,
\author{Carlos Mejuto-Zaera}{
\UCB
}%
,
\author{Xiao Mi}{
\Google
}%
,
\author{Masoud Mohseni}{
\Google
}%
,
\author{Wojciech Mruczkiewicz}{
\Google
}%
,
\author{Josh Mutus}{
\Google
}%
,
\author{Ofer Naaman}{
\Google
}%
,
\author{Matthew Neeley}{
\Google
}%
,
\author{Charles Neill}{
\Google
}%
,
\author{Hartmut Neven}{
\Google
}%
,
\author{Michael Newman}{
\Google
}%
,
\author{Murphy Yuezhen Niu}{
\Google
}%
,
\author{Thomas E.~O'Brien}{
\Google
}%
,
\author{Eric Ostby}{
\Google
}%
,
\author{B\'{a}lint Pat\'{o}}{
\Google
}%
,
\author{Andre Petukhov}{
\Google
}%
,
\author{Harald Putterman}{
\Google
}%
,
\author{Chris Quintana}{
\Google
}%
,
\author{Jan-Michael Reiner}{
\HQS
}%
,
\author{Pedram Roushan}{
\Google
}%
,
\author{Nicholas C.~Rubin}{
\Google
}%
,
\author{Daniel Sank}{
\Google
}%
,
\author{Kevin J.~Satzinger}{
\Google
}%
,
\author{Vadim Smelyanskiy}{
\Google
}%
,
\author{Doug Strain}{
\Google
}%
,
\author{Kevin J.~Sung}{
\Google,\!
\UM
}%
,
\author{Peter Schmitteckert}{
\HQS
}%
,
\author{Marco Szalay}{
\Google
}%
,
\author{Norm M.~Tubman}{
\NASA
}%
,
\author{Amit Vainsencher}{
\Google
}%
,
\author{Theodore White}{
\Google
}%
,
\author{Nicolas Vogt}{
\HQS
}%
,
\author{Z.~Jamie Yao}{
\Google
}%
,
\author{Ping Yeh}{
\Google
}%
,
\author{Adam Zalcman}{
\Google
}%
,
\author{Sebastian Zanker}{
\HQS
}%

\bigskip

{\xGoogle}
{\xUMass}
{\xCaltech}
{\xUCSB}
{\xNASA}
{\xKBR}
{\xBosch}
{\xSherbrooke}
{\xUCR}
{\xHQS}
{\xOxford}
{\xUCB}
{\xUM}

\end{flushleft}

\twocolumngrid

\putbib[fhm]
\end{bibunit}

\clearpage
\renewcommand\appendixname{}
\renewcommand\appendixpagename{\Large{Supplementary Information}}
\makeatletter
\def\p@subsection{}
\makeatother
\renewcommand*{\thesubsection}{\Alph{section}\arabic{subsection}}
\renewcommand*{\thesubsubsection}{}
\renewcommand\thefigure{S\arabic{figure}}    
\setcounter{figure}{0}
\setcounter{equation}{0}
\makeatletter
\newcommand*{\newbibstartnumber}[1]{%
  \apptocmd{\thebibliography}{%
    \global\c@NAT@ctr #1\relax
    \addtocounter{NAT@ctr}{-1}%
  }{}{}%
}
\makeatother
\newbibstartnumber{43}

\begin{appendices}

\begin{bibunit}[apsrev4-1_with_title]

\section{Gate decompositions}
\label{sec:gate_decompositions}

\begin{figure*}[ht]
\centering
    \includegraphics[width=0.9\textwidth]{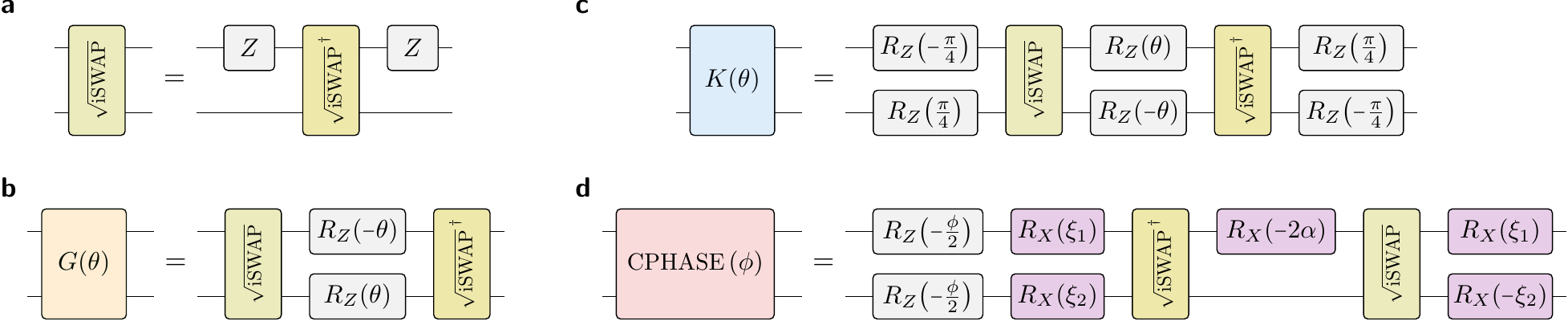}
    \caption{\textbf{Gate decomposition.} Decomposition of various two-qubit gates into single-qubit gates and two $\sqrt{\iSWAP}^{\,\dag} \equiv \KK(\pi/4)$ gates, an approximate description of our native two-qubit gate.  {\textbf a.} The two gates $\sqrt{\iSWAP}$ and its inverse (Hermitian conjugate) are equivalent up to single-qubit Pauli-$Z$ gates. {\textbf b.} Decomposition of the Givens rotation $G(\theta)$ for initial state preparations, where $R_{Z}(\theta) = \exp(\mathord -i\theta Z/2)$. {\textbf c.} Decomposition of the $\KK(\theta)= e^{-i\theta (X\otimes X+Y\otimes Y)/2}$ gate. {\textbf d.} Decomposition of the $\CPHASE\, (\phi)=\diag(1,1,1,e^{-i\phi})$ gate, where $R_{X}(\xi) = \exp(\mathord-i\xi X  / 2)$ and the parameters $\xi_1$, $\xi_2$, and $\alpha$ are functions of $\phi$. }
   \label{fig:sqrt_decompositions}
\end{figure*}

In Fig.~\hyperref[{fig:sqrt_decompositions}]{\ref{fig:sqrt_decompositions}}, we provide decompositions of the parameterized two-qubit gates used in this work into the standard $\sqrt{\iSWAP}^{\,\dag}$ gate, i.e., $\KK(\pi/4)$.  Our native two-qubit gate is close to $\KK(\pi/4)$ and can be better described by $\KK(\vartheta)\,\CPHASE(\varphi)$, where $\vartheta\approx \pi/4$ and  $\varphi\approx \pi/23$.  Here we provide a derivation on decomposing the \CPHASE\ gate into two native two-qubit gates and several single-qubit gates.  To simplify notations, we introduce the gate
\begin{align}\label{eq:Sycamore_b}
    \FF(\vartheta, \varphi)&= e^{-i\vartheta(X\otimes X + Y\otimes Y)/2-i\varphi Z\otimes Z/4}\,,
\end{align}
which is related to our native gate by $Z$ rotations
\begin{align}
    \KK(\vartheta)\,\CPHASE(\varphi) = e^{-i\varphi/4}\, e^{i\varphi (Z_1 + Z_2)/4}\:\FF(\vartheta, \varphi) \,,
\end{align}
where $Z_1 = Z\otimes I$ and $Z_2 = I\otimes Z$.  The sign of $\vartheta$ can be flipped by using single-qubit $Z$ gates,
\begin{align}
  Z_1\, \FF(\vartheta, \varphi)\, Z_1  = Z_2\, \FF(\vartheta, \varphi)\, Z_2  = \FF(-\vartheta, \varphi) \,.
\end{align}
Sandwiching a microwave gate with two two-qubit gates of opposite $\vartheta$, we have 
\begin{align}\label{eq:composite_gate_schmidt}
  \FF(-\vartheta, \varphi)\, e^{i\alpha X_1}\xssp \FF(\vartheta, \varphi) =  \Gamma_1\otimes I -iZ\otimes \Gamma_2\,.
\end{align}
where $X_1=X\otimes I$ and the Schmidt operators are
\begin{align}
    &\Gamma_1(\alpha) = \cos\alpha\cos(\varphi/2)\, I + i\sin\alpha\cos\vartheta\, X\,,\\[3pt]
    &\Gamma_2(\alpha) = \cos\alpha\sin(\varphi/2)\, Z - \sin\alpha\sin\vartheta\, Y\,,
\end{align}
with the Schmidt coefficients $\opnorm{\Gamma_1}$ and $\opnorm{\Gamma_2}$.  The unitary~\eqref{eq:composite_gate_schmidt} is equivalent to a \CPHASE\ gate up to single-qubit gates, 
\begin{align}
\CPHASE(\phi) = e^{-i\phi(I-Z)\otimes (I-Z)/4} \,,
\end{align}
which has two non-zero Schmidt coefficients  $\norm{\cos(\phi/4)}$ and $\norm{\sin(\phi/4)}$.  We require $\opnorm{\Gamma_2(\alpha)} = \norm{\sin(\phi/4)}$ to match the Schmidt coefficients of the two unitaries, which yields 
\begin{align}\label{eq:alpha}
\sin\alpha = \sqrt{\frac{\sin(\phi/4)^2 - \sin(\varphi/2)^2}{\sin(\vartheta)^2-\sin(\varphi/2)^2}}\;.
\end{align}
This equation can be solved when one of the following two conditions is satisfied
\begin{align}
\lvert\sin\vartheta\rvert \leq \norm{\sin(\phi/4)} \leq \lvert\sin(\varphi/2)\rvert\,,\\[3pt]
\lvert\sin(\varphi/2)\rvert\leq \norm{\sin(\phi/4)}  \leq \lvert\sin\vartheta\rvert \,.
\end{align}
For the parameters in our native gate, these conditions can be simplified to $\norm{\phi} \geq 2\xssp \norm{\varphi}$.  To match the Schmidt operators on the first qubit, we introduce two $X$ rotations with the same angle
\begin{gather}
 R_{\nsp X}(\xi_1)\, \Gamma_1(\alpha)\, R_{\nsp X}(\xi_1) = \cos(\phi/4)\,I\,,\\[2pt]
  R_{\nsp X}(\xi_1)\, Z\,  R_{\nsp X}(\xi_1)=  Z\,,
\end{gather}
where  $R_{\nsp X}(\xi) = e^{-i\xi X / 2}$ and
\begin{align}
 \xi_1 = \tan^{-1} \bigg(\frac{\tan\alpha \cos\vartheta}{\cos(\varphi/2)}\bigg)
 + \frac{\pi}{2} \Bigl(\nsp 1-\sgn\Bigl(\nsp\nsp\cos\frac{\varphi}{2}\Bigr)\Bigr)\,.
\end{align}
To match the Schmidt operators on the second qubit, we introduce two $X$ rotations with opposite angles
\begin{align}
 R_{\nsp X}(\mathord -\xi_2)\, \Gamma_2(\alpha)\, R_{\nsp X}(\xi_2)= \sin(\phi/4)\,Z\,,
\end{align}
where 
\begin{align}
\xi_2 = \tan^{-1} \bigg(\frac{\tan\alpha \sin\vartheta}{\sin(\varphi/2)}\bigg) + \frac{\pi}{2} \Bigl(\nsp 1-\sgn\Bigl(\nsp\nsp\sin\frac{\varphi}{2}\Bigr)\Bigr)\,.
\end{align}
Put everything together, we have
\begin{align}
 & R_{\nsp X}(\xi_1,-\xi_2)\,\FF(-\vartheta, \varphi)\, e^{i\alpha X_1}\, \FF(\vartheta, \varphi)\, R_{\nsp X}(\xi_1, \xi_2)\nonumber\\[2pt]
 & \qquad= \cos(\phi/4)\: I \otimes I - i\sin(\phi/4)\, Z\otimes Z\nonumber\\[2pt]
 & \qquad= e^{i\phi/4}\, e^{-i\phi (Z_1 + Z_2)/4}\, \CPHASE(\phi)\,, \label{eq:final_decomposition}
\end{align}
where $R_{\nsp X}(\xi_1, \xi_2) = R_{X}(\xi_1)\otimes R_{X}(\xi_2)$.  This implements the desired \CPHASE\ gate up to single-qubit $Z$ rotations.

\section{spin-charge separation by Bosonization}
\label{sec:spin_charge_boson}

The bosonization theory discussed here only applies to low-energy and long-wavelength excitations in the 1D Fermi-Hubbard model, whereas the quenched dynamics presented in the main text involves highly excited states with short wavelengths.  As a result, the theory can only be used to explain the findings in the main text qualitatively. 

The dispersion relation of a particle in a 1D homogeneous quantum liquid is linear $\epsilon_k = \pm v_f k$, and both the spin and charge excitations travel at the Fermi velocity $v_c = v_s = v_f$.  For nonzero interaction strengths, the spin and charge-wave packages propagate at different velocities.  As was proposed by Haldane~\cite{haldane_luttinger_1981, Fermi_Luttinger_Liquids2000}, the 1D spin-1/2 Fermi gas can be mapped to an effective hydrodynamic Hamiltonian, which describes the original system faithfully at wavelengths much larger than the interparticle spacing. This theory of noninteracting bosons is called Luttinger liquid, where all correlation functions can be exactly calculated. This Hamiltonian takes the form
\begin{align}
 H = \sum_{\alpha = c, s} \int \dif x\; \frac{\hbar v_\alpha}{2}\left[K_{\alpha} \Pi_{\alpha}^2 + \frac{1}{K_{\alpha}}\big(\partial_x\phi_\alpha\big)^2\right]\,,
\end{align}
where $\phi_\alpha$ is a bosonic field operator and $ \Pi_{\alpha}$ its conjugate momentum operator.  The low-energy physics is completely characterized by the phenomenological parameters: the density-wave velocity $v_\alpha$ and Luttinger parameter $K_{\alpha}$, which depends on the interaction \cite{Bosonization_Luttinger_Liquid_1995}. The single particle spectral function of 1D Fermi liquid has two power-law singularities for the spin and charge excitations respectively~\cite{zacher_systematic_1998,benthien_spectral_2004}.

\section{Floquet calibration}
\label{sec:floq_calib}

Calibration of quantum gates is one of the most crucial steps in achieving high-fidelity quantum computation and its large-scale deployment~\cite{erhard_characterizing_2019, kelly_physical_2018, klimov_snake_2020}.  Temporal instabilities, including drifts and fluctuations in the control fields and qubit frequencies~\cite{bylander_noise_2011, megrant_planar_2012, fogarty_nonexponential_2015, klimov_fluctuations_2018, chan_assessment_2018, wan_quantum_2019, burnett_decoherence_2019, proctor_detecting_2019}, can propagate and accumulate coherently in large quantum circuits.  Therefore, it is crucial to develop fast and accurate calibration methods to characterize and mitigate these errors.  However, common calibration tools, such as randomized benchmarking~\cite{knill_randomized_2008,magesan_scalable_2011}, compressed sensing~\cite{shabani_efficient_2011, magesan_compressing_2013}, gate set tomography~\cite{greenbaum_introduction_2015, blume-kohout_demonstration_2017}, and cross-entropy benchmarking~\cite{boixo_characterizing_2018} are often too slow to capture drifts and fluctuations in the hardware. 

Quantum metrology~\cite{braunstein_generalized_1996, giovannetti_advances_2011} offers a route to this goal.  The Heisenberg limit $\mathcal{O}(1/n)$ sets a fundamental lower error bound in phase estimation with $n$ photons~\cite{caves_quantum-mechanical_1981, yurke_su2_1986, holland_interferometric_1993, lee_quantum_2002}, whereas the standard quantum limit $\mathcal{O}(1/\sqrt n)$ refers to the minimum uncertainty allowed by using semi-classical states.  Modified versions of the quantum phase-estimation algorithm~\cite{kitaev_quantum_1995, cleve_quantum_1998, nielsen_quantum_2002} have been shown to reach Heisenberg scaling theoretically~\cite{summy_phase_1990, luis_optimum_1996, wiseman_adaptive_1997, berry_optimal_2000, de_burgh_quantum_2005, boixo_parameter_2008} and experimentally~\cite{higgins_entanglement-free_2007, higgins_demonstrating_2009}.  Based on this idea, Kimmel~\textit{et al}.~\cite{kimmel_robust_2015, rudinger_experimental_2017} proposed a protocol to characterize the axis and angle of a single-qubit rotation.  It achieves uncertainty $\mathcal O(1/n)$ by repeating identical operations $\mathcal O(n)$ times. 

Here we describe a fast and accurate calibration protocol for entangling gates.  It is based on the idea that an entangling gate can be uniquely determined by the eigenvalues of the composite gates making up the entangling gate and different sets of single-qubit gates. By repeating the composite gate $n$ times in a quantum circuit, small changes in the gate parameters are amplified $n$ times.  Repeating the cycle unitary for many times also makes the protocol robust to state preparation and measurement (SPAM) errors.
In these respects, our protocol resembles gate set tomography, with the composite gates playing the role of ``germs'' therein, though by prioritizing the errors we wish to calibrate and leveraging well the form of single-qubit gates we require far fewer resources than is typical for gate set tomography.  We show that this protocol can be implemented both adaptively and non-adaptively.  In Sec.~\ref{subsec:n_conserv}, we introduce the procedure of characterizing two-qubit gates that conserves the numbers of excitations.  In Subsec.~\ref{sec:floq_calib:phase_estimation_decoherence}, we discuss the effects of decoherence on the ultimate precision of the procedure.  In Subsec.~\ref{sec:floq_clib:cyc_rep_num}, we study how to best choose the cycle repetition numbers.   In Subsec.~\ref{sec:floq_calib:general}, we show that in principle our procedure can be applied to general multi-qubit gates.

\subsection{Excitation-number-conserving gates}
\label{subsec:n_conserv}

\begin{figure}
\label{fig:floquet_angles}
\centering
   \includegraphics[width=0.48\textwidth]{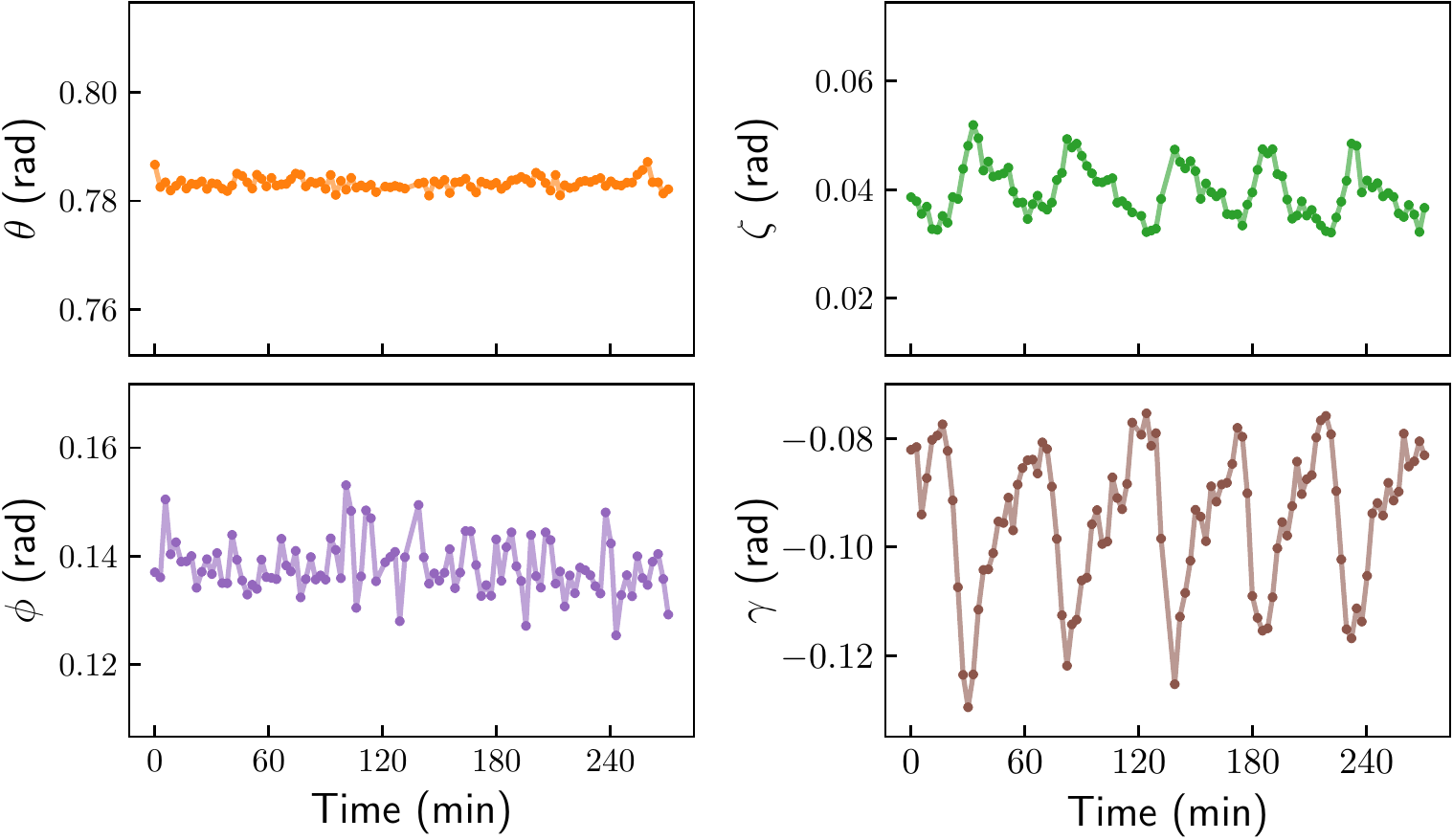}
    \caption{The values of the four (out of five) parameters  of our hardware native gate obtained by Floquet calibration.  The mechanism of the oscillations in the parameters $\zeta$ and $\gamma$ is still not completely clear, but likely due to fluctuations in temperature.  Since the changes of these two parameters are significant, they must be corrected in real time to obtain desired results. }
\end{figure}

The most general excitation-number-conserving two-qubit gate takes the following form with the basis states in the order $\ket{00}$, $\ket{01}$, $\ket{10}$, and $\ket{11}$,
\begin{align}\label{eq:nc_2q_u}
&\hspace{-0.7em}\NCG(\theta, \zeta, \chi, \gamma, \phi) = \nonumber\\[3pt]
&\hspace{-0.7em}
\scalebox{0.97}{$
\begin{pmatrix}
1    &0    &0   &0\\[3pt]
0 &e^{-i(\gamma + \zeta)} \cos\theta &  -i\, e^{-i(\gamma-\chi)}\sin \theta & 0\\[3pt]
0 &-i\, e^{-i(\gamma+\chi)}\sin \theta &e^{-i(\gamma - \zeta)} \cos\theta& 0\\[3pt]
0 &0 &0 &e^{-i(2\gamma + \phi)}
\end{pmatrix}$}\,,
\end{align}
where $0\leq\theta\leq \pi/2$ is the \iSWAP\ angle, $\phi$ is the controlled-phase angle, and $\zeta$, $\chi$, and $\gamma$ are single-qubit phase angles.  In Fig.~\ref{fig:floquet_angles}, we plot the values of the parameters of our hardware native gate (except for $\chi$) obtained by Floquet calibration over a time period of several hours.  We denote the single-qubit $Z$ rotation as $R_Z(\zzz) = \diag(1,\, e^{i\zzz})$, equivalent to the definition $R_Z(\zzz) =\diag(e^{-i\zzz/2},\, e^{i\zzz/2})$ in Cirq~\cite{cirq} up to an overall phase.  Single-qubit $Z$ rotations acting on two qubits take the form
\begin{align}
R_Z(\zzz_1, \zzz_2) &= \diag\pigl(1, e^{i\zzz_2}, e^{i\zzz_1}, e^{i(\zzz_1 +\zzz_2)}\pigr)\\
& = \NCG(0, \zzz^\mym, 0,-\zzz^\myp, 0)\,,
\end{align}
where $\zzz^\mypm = (\zzz_1\pm\zzz_2)/2$.  The general number-conserving gate  defined in Eq.~\eqref{eq:nc_2q_u} can be decomposed into the sequence
\begin{align}\label{eq:gfsim_fsim}
&\hspace{-0.4em} \NCG(\theta, \zeta, \chi, \gamma, \phi)=\nonumber\\[3pt]
&\: R_Z(-\gamma, -\gamma)\, R_Z(\beta, -\beta)\, \NCG(\theta, 0, 0, 0, \phi) \, R_Z(\alpha, -\alpha)\,,
\end{align}
where $\alpha = (\zeta+\chi)/2$, $\beta = (\zeta-\chi)/2$.  
It also takes the block diagonalized form,
\begin{align}
 \NCG=\diag\Bigl(1,\: e^{-i\gamma} u(\theta,\zeta,\chi),\: e^{-i(2\gamma + \phi)}\Bigr)\,,
\end{align}
where the $2\times 2$ matrix $u$ reads
\begin{align}\label{eq:nc_2q_u_reduced}
u(\theta, \zeta, \chi) &= 
\begin{pmatrix}
e^{-i\zeta} \cos\theta &  -i\, e^{i\chi}\sin \theta \\[3pt]
-i\, e^{-i\chi}\sin \theta &e^{i \zeta} \cos\theta
\end{pmatrix}\\[3pt]
&= I\cos\Omega(\theta, \zeta) -i\, \sigma(\theta, \zeta, \chi) \sin\Omega (\theta, \zeta) \label{eq:pauli_rep}\,,
\end{align}
where $\Omega(\theta, \zeta) = \arccos(\cos\theta\cos\zeta) \in [0, \pi]$ is the Rabi angle and the idempotent matrix $\sigma$ reads
\begin{align}
&\hspace{-0.4em}\sigma(\theta, \zeta, \chi)=\nonumber\\[2pt]
&\hspace{0.7em}  \pigl((X\cos\chi - Y \sin\chi) \sin\theta + Z \cos\theta\sin\zeta\pigr) \big/ \sin \Omega\,.
\end{align}
The eigenstates of $\sigma(\theta, \zeta, \chi)$ with eigenvalues $\pm 1$ are 
\begin{align}
\ket{\psi^+} &= 
    \cos(s /2)\,\ket{0} + \sin(s /2)\,e^{-i\chi}\,\ket{1}\,,\\[2pt]
    \ket{\psi^-} &= 
    \sin(s /2)\,\ket{0} - \cos(s /2)\,e^{-i\chi}\,\ket{1}\,,
\end{align}
where $s = \arccot (\cot \theta \sin\zeta)\in [0, \pi]$.  
The $\nn$-th power of the number-conserving gate reads
\begin{align}
\NCG^n = \diag\pigl(1,\: e^{-in\gamma} u(\theta,\zeta,\chi)^n,\: e^{-in(2\gamma + \phi)}\pigr)\,,
\end{align}
where $u^\nn$ can be solved using the representation~\eqref{eq:pauli_rep},
\begin{align}
u^\nn
&= I \cos(\nn\xssp\Omega) - i\, \sigma(\theta, \zeta, \chi) \sin(\nn\xssp \Omega)\,,
\end{align}
which takes the matrix form
\begin{align}
\hspace{-0.3em}
\scalebox{0.92}{
$\begin{pmatrix}
\cos(\nn\xssp\Omega)-i\Lambda_n\cos\theta\sin\zeta &  -i\xssp \Lambda_n\xssp e^{i\chi}\sin \theta \\[4pt]
-i\xssp \Lambda_n\xssp e^{-i\chi}\sin \theta &\cos(\nn\xssp\Omega)+i\Lambda_n\cos\theta\sin\zeta
\end{pmatrix}$}\,,\label{eq:un}
\end{align} 
where $\Lambda_n = \sin(\nn\xssp \Omega)/\sin \Omega$.

To calibrate the gate parameters, we introduce the cycle unitary made up of $U$ and singe-qubit $Z$ rotations
\begin{align}\label{eq:cycle}
  U_c &\equiv  \NCG(\theta, \zeta, \chi, \gamma, \phi)\,R_Z(\zzz_1, \zzz_2)\\[2pt]
  &= \NCG(\theta, \zeta_c, \chi_c, \gamma_c, \phi)\,,
\end{align}
where the parameters of $U_c$ are related to the original ones via the linear relations
\begin{align}\label{eq:trans_rule}
\zeta_c = \zeta + \zzz^\mym\,,\quad \chi_c = \chi + \zzz^\mym\,,\quad \gamma_c =  \gamma - \zzz^\myp\,.
\end{align}
The two parameters $\zzz^\mypm = (\zzz_1 \pm \zzz_2) /2$ can be controlled by adjusting the single-qubit pulses.  The hidden assumption here is: the two-qubit gate $\NCG$ does not depend on $\zzz_1$ and $\zzz_2$, i.e., the control pulses do not interleave (no gate bleeding).  The cycle unitary has two trivial eigenstates $\ket{00}$ and $\ket{11}$, 
\begin{figure}[t]
\label{fig:omega_z}
\centering
    \includegraphics[width=0.3242174629324546\textwidth]{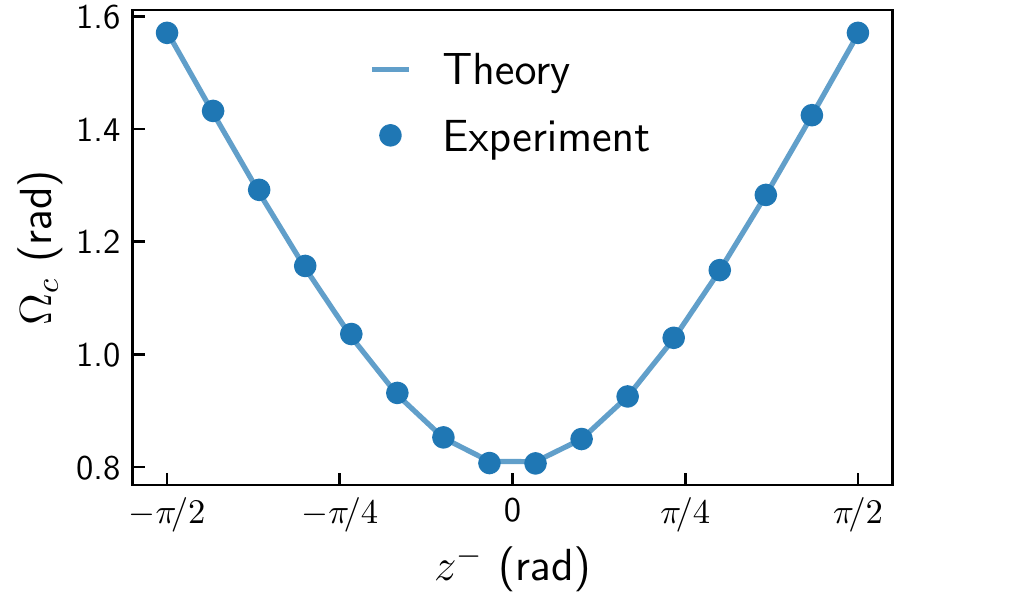}
    \caption[]{Experiment data for $\Omega_c$ as a function of $\zzz^\mym$.  They compare well with  the fitted curve based on the analytic expression~\eqref{eq:omega_eta}.  This indicates that gate bleeding between $R_Z$ and the two-qubit gate $U$ is negligible, i.e., their control pulses do not interleave with each other.}
\end{figure}

\begin{align}
U_c \ket{00} = \ket{00}\,,\quad U_c \ket{11} = e^{-i(2\gamma_c+\phi)}\ket{11}
\end{align}
and two nontrivial eigenstates
\begin{align}
\ket{\Psi_c^+} &= 
    \cos(s_c /2)\,\ket{01} + \sin(s_c /2)\,e^{-i\chi_c}\,\ket{10}\,,\\[2pt]
    \ket{\Psi_c^-} &= 
    \sin(s_c /2)\,\ket{01} - \cos(s_c /2)\,e^{-i\chi_c}\,\ket{10}\,,
\end{align}
where $s_c = \arccot (\cot \theta \sin\zeta_c)$.  The corresponding eigenvalue equations are
\begin{align}\label{eq:cycle_Omega}
 U_c \ket{\Psi^\pm_c} = e^{-i (\gamma_c \pm \Omega_c)}\xssp \ket{\Psi^\pm_c}\,,
\end{align}
where $\Omega_c \in [\theta, \pi-\theta]$ is the Rabi angle 
\begin{align}\label{eq:omega_eta}
\Omega_c =
\arccos\bigl(\nsp\cos\theta\cos(\zeta + \zzz^\mym)\bigr)\,.
\end{align}
Knowing the eigenvalues of $U_c$ allows one to learn $\gamma_c$, $\Omega_c$, and $\phi$, from which one can learn $\gamma$ using the last identity in Eq.~\eqref{eq:trans_rule}.  To learn $\theta$ and $\zeta$ using Eq.~\eqref{eq:omega_eta}, one needs to know $\Omega_c$ for at least two different values of $\zzz^\mym$.  In Fig.~\ref{fig:omega_z}, we plot the experiment results for $\Omega_c$ as a function of $\zzz^\mym$ and compare them with the fitted curve based on Eq.~\eqref{eq:omega_eta}.  They conform extremely well with each other, indicating that gate bleeding between the single-qubit $Z$ rotations and the two-qubit $U$ is negligible.

In the following, we introduce three sets of calibration circuits to learn the five parameters in the number-conserving gate~\eqref{eq:nc_2q_u}.  We run each set of circuits with the cycle repetition numbers from the set  
\begin{align}\label{eq:set_of_n}
\mathbb N = \bigl\{\lceil r^{k} \rceil \,\big\lvert\, k = 0,1,\ldots, K-1\bigr\}\,,\quad r>0\,.
\end{align}
This is necessary because the eigenvalues of $U_c$ can only be determined up to modulo $2\pi/n$ when $U_c$ is repeated for $\nn$ times.  To make sure that we search in the correct principal region, the true value should be located within the $(\pi/\nn)$-neighborhood of the prior estimate with high probability.  We will discuss this issue and how to choose the real number $r$ in more detail in Section~\ref{sec:floq_calib:phase_estimation_decoherence}.

\subsubsection{Calibration circuits 1}
\label{subsubsec:ins_1}

\begin{figure}[tp]
\label{fig:floquet_fsim_ins0}
\centering
   \includegraphics[width=0.3541844\textwidth]{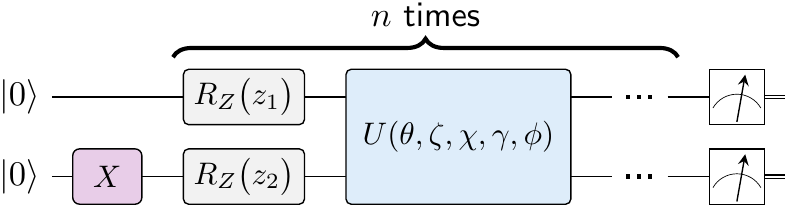}
    \caption{The circuit to calibrate 
    the parameters $\theta$ and $\zeta$ in the excitation-number-conserving gate.  The cycle unitary is repeated for $\nn$ times before we measure the qubits in the $Z$ basis.}
\end{figure}

This set of calibration circuits are used to learn the parameters $\theta$ and $\zeta$ in the gate~\eqref{eq:nc_2q_u}, see Fig.~\ref{fig:floquet_fsim_ins0}.  The probability of measuring the state $\ket{10}$ is
\begin{align}\label{eq:est_Omega}
   p_{\nn} &= \normb{\bra{10}\, U_c^\nn R_{\nsp X}(0,\pi) \ket{00}}^2\nonumber\\[2pt]
   &= \normb{\bra{1}  u_c^\nn \ket{0}}^2 = \pigl(
   \sin(\nn\xssp \Omega_c) \sin\theta / \sin \Omega_c\pigr)^2\,,
\end{align}
where 
$u_c^n\equiv u(\theta,\zeta_c,\chi_c)^n$ is given in Eq.~\eqref{eq:un}.
By postselecting the measurement results with one excitation, i.e., $\ket{01}$ and $\ket{10}$, we make the results robust to $T_1$ error and bit-flip errors in the measurement.  An unbiased estimator of $p_{\nn}$ based on the measurement results is
\begin{align}
   \hat p_{\nn} =  \frac{\text{Number of outcome 10}}{\text{Number of outcomes 01 and 10}}\,,
\end{align}
and its variance decreases as the number of measurements passing postselection (the denominator) increases. 

Using Eqs.~\eqref{eq:omega_eta} and \eqref{eq:est_Omega}, we can relate the measurement probability $p_n$ to the gate parameters $\theta$ and $\zeta$ and the adjustable variable $\zzz^\mym$.  For $\nn=1, 2$, we have
\begin{align}
 \norm{\sin\theta} = \sqrt{p_{1}}\,,\quad   \sin(2\theta)\,\norm{\cos(\zeta + \zzz^\mym)} = \sqrt{p_{2}}\,,
\end{align}
which can be used to get initial estimates of $\theta$ and $\zeta$.  To get robust estimates for larger $\nn$, we use aggregated results from several previous runs by introducing the cost function (we use the Hellinger distance, but other metrics might work as well)
\begin{align}\label{eq:cost_circuit_1}
 &\hspace{-0.3em}C_\nn^\ell(x, y) \nonumber\\
 &\hspace{0.4em} = \!\sum_{\substack{\ell \leq \mm \leq \nn\\[1pt] \mm\in \mathbb N}}\!\sum_{\;\;\zzz^\mym \in\mathbb Z_m^-} \biggl(
   \Bigl\lvert\frac{\sin(\mm\xssp \Omega_c) \sin x}{\sin \Omega_c}\Bigr\rvert - \sqrt{\hat p_{\mm,\xssp\zzz^\mym}}\biggr)^2\,,
\end{align}
where $\Omega_c(x,y,\zzz^\mym) = \arccos\bigl(\cos x\cos(y + \zzz^\mym)\bigr)$ and the lower bound $\ell$ determines the number of terms included in the cost function.  This approach also provides us with the flexibility of using a different set of $\zzz^\mym$ for each cycle repetition number $\mm$, which we denote as $\mathbb Z_\mm^-$.  When $\nn$ is small, we use the global minimum of the cost function as the estimators of $\theta$ and $\zeta$.  For larger $\nn$, the landscape of the cost function becomes rugged, and we use its local minimum around the prior estimates as the new estimates.  We choose $\ell$ based on the following rules.  For small $\nn$, it is easy to get into the wrong principal regions, and we include all the prior runs in the cost function by setting $\ell = 1$.  As $\nn$ increases, we fix the number of terms in the cost function by increasing $\ell$.  We then gradually reduce the number of terms in the cost function to minimize the variance of the estimates at the end. 

The variances of the estimates of $\theta$ and $\zeta$ diverge when either $\norm{\partial \Omega_c/ \partial \theta}$ or $\norm{\partial \Omega_c/ \partial \zeta}$ is close to zero.  This can be avoided by adaptively choosing the values of $\zzz^\mym$ based on the current estimates of $\theta$ and $\zeta$.   We may choose the two values of $\zzz^\mym$ close to $\pi/4$ and $3\pi/4$, which are apart by $\pi/2$ for best performance.  This choice also leaves a large margin between $\zeta_c\; (\mathord{\bmod}\; \pi/2)$ and $0$ for small $\zeta$, where either $\norm{\partial \Omega_c/ \partial \theta}$ or $\norm{\partial \Omega_c/ \partial \zeta}$ equals to zero.  The standard error in the estimate of $\Omega_c$ is inversely proportional to the derivative
\begin{align}\label{eq:dp_domega}
  \frac{\partial p_{\nn}}{\partial \Omega_c} \simeq \nn\sin(2\nn\xssp \Omega_c)\, \bigl(\sin\theta / \sin \Omega_c\bigr)^2\,,
\end{align}
where we assume $\nn\gg 1$ and neglect terms of order $\mathcal O(1)$.  We maximize the fast oscillating part $\sin(2\nn\xssp \Omega_c)$ in Eq.\eqref{eq:dp_domega} by choosing $\zzz^\mym$ from the neighborhood of $\pi/4$ and $3\pi/4$ such that
\begin{align}\label{eq:eta_condition_2}
    \nn\, \Omega_c  = \pi / 4 \;(\mathord{\bmod}\; \pi/2)\,,
\end{align}
where $\Omega_c$ is evaluated using the estimates of $\theta$ and $\zeta$.  

These calibration circuits can also be implemented non-adaptively at the price of increasing the number of circuits.  We run the circuit in Fig.~\ref{fig:floquet_fsim_ins0} for several equidistant values of $\zzz^\mym$ in the two intervals $\pi/4\pm w / \nn$ and $3\pi/4\pm w / \nn$, where $w =  \pi/(2 \cos\theta)$.  This choice guarantees that Eq.~\eqref{eq:dp_domega} has big values at least for some of the selected values of $\zzz^\mym$.  The values of $\theta$ and $\zeta$ can then be estimated by fitting the data to Eq.~\eqref{eq:est_Omega}.

\subsubsection{Calibration circuits 2}

\begin{figure}[tp]
\label{fig:floquet_fsim_ins1}
\centering
   \includegraphics[width=0.41347518\textwidth]{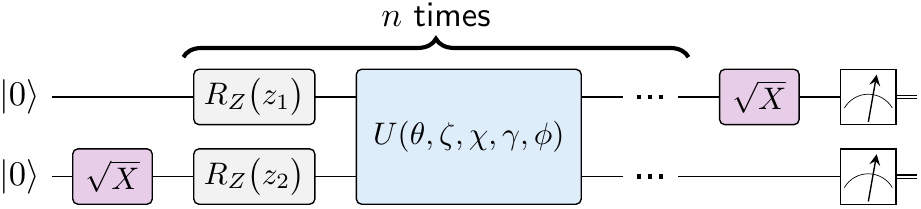}
    \caption{The circuit to calibrate 
    the parameters $\gamma$ and $\chi$ in the general \FSIM\ gate.  The initial $\sqrt X$ gate creates a superposition between the state $\ket{00}$ and the single-excitation state $\ket{01}$.}
\end{figure}

Here we will focus on $\gamma$ and $\chi$, since $\theta$ and $\zeta$ can be better estimated with the former calibration circuits.  The circuit in Fig.~\ref{fig:floquet_fsim_ins1} can be used to measure the accumulated phase $\nn\gamma + \chi$, from which one can estimate $\gamma$ with high precision.  In comparison, $\chi$ can only be estimated with low precision (also susceptible to SPAM errors) due to lack of the scaling factor $\nn$.  Consider the measurement probabilities for the circuit in Fig.~\ref{fig:floquet_fsim_ins1},
\begin{align}
\label{eq:ins1_p}
    p_\nn &= \normb{\bra{00} R_{\nsp X}(\pi/2,0)\, U_c^\nn R_{\nsp X}(0,\pi/2)\, \ket{00}}^2\nonumber\\[2pt]
    &= \frac{1}{4}\, \normb{e^{i\nn \gamma_c} -  \bra{1} \, u_c^\nn  \ket{0}}^2\,,
    \end{align}
    and
    \begin{align}
    q_\nn  &= \normb{\bra{01} R_{\nsp X}(\pi/2,0)\, U_c^\nn R_{\nsp X}(0,\pi/2)\, \ket{00}}^2 \nonumber\\[2pt]
    &= \frac{1}{4}\, \norm{\bra{0} \, u_c^\nn  \ket{0}}^2\,. \label{eq:nun00}
\end{align}
To estimate the parameter $\gamma_c=\gamma-\zzz^\myp$ using Eq.~\eqref{eq:ins1_p}, we will need to learn $ \bra{1} u_c^\nn \ket{0} $.  Its phase can be calculated using Eq.~\eqref{eq:un},
\begin{align}
    \arg(\bra{1} \, u_c^\nn  \ket{0}) 
    &= -\chi - \zzz^\mym -\frac{\pi}{2}\sgn\Lambda_\nn\,,
\end{align}
where $\Lambda_n = \sin(\nn\xssp \Omega_c)/\sin \Omega_c$ and its sign can be determined with confidence provided that we have good estimates of $\theta$ and $\zeta$.  Using the normalization condition $\norm{\bra{0} \, u_c^\nn  \ket{0}}^2 + \norm{\bra{1} \, u_c^\nn  \ket{0}}^2 = 1$, we have
\begin{align}
   \norm{\bra{1} \, u_c^\nn  \ket{0}} = \sqrt{1-4q_\nn} \,.
\end{align}
The measurement probability $p_\nn$ in Eq.~\eqref{eq:ins1_p} is a function of $\norm{\bra{1} \, u_c^\nn  \ket{0}}$ and the relative phase
\begin{align}
   \mu_\nn &=  \nn \gamma_c - \arg(\bra{1} \, u_c^\nn  \ket{0})\\
   &= (\nn \gamma +\chi) - \nn\zzz^\myp + \zzz^\mym +\frac{\pi}{2}\sgn\Lambda_\nn\,.\label{eq:mu_gamma_phase}
\end{align}
Knowing $\mu_\nn$ for different values of $\nn$ allows one to estimate $\gamma$ and $\chi$.  It is related to the measurement probabilities through the relation
\begin{align}
    &\normb{e^{i\mu_\nn} - \sqrt{1-4q_\nn}\:}^2 = 4p_\nn\,,
\end{align}    
or equivalently
\begin{align}\label{eq:amp_to_prob}
  \cos \mu_\nn = \frac{1- 2( p_\nn+q_\nn)}{\sqrt{1-4q_\nn}}\,.
\end{align}
Again, we introduce the cost function
\begin{align}
  C_\nn^\ell(x, y) 
 &= \sum_{\substack{\ell \leq \mm \leq \nn\\[1pt] \mm\in \mathbb N}} \biggl(
    f_\nn(x,y) -  \frac{1- 2(\hat p_\nn+\hat q_\nn)}{\sqrt{1-4\hat q_\nn}}\,\biggr)^2\,,
\end{align}
where $f_\nn(x,y) = \cos(\nn x +y - \nn\zzz^\myp + \zzz^\mym +\frac{\pi}{2}\sgn\Lambda_\nn)$.  The parameters $\gamma$ and $\chi$ can be estimated by using the arguments $x$ and $y$ that minimize the cost function, respectively.  We follow the same prescription as in the former case to choose $\ell$ and minimize the cost function.

To reduce the variances of estimators, we choose $\zzz^\mym$ such that $\norm{\bra{1} u_c^\nn \ket{0}}$ is maximized using the estimates of $\theta$ and $\zeta$.  We also choose $\zzz^\myp$ such that $\mu_\nn \simeq \pi/2\; (\mathord{\bmod}\; \pi)$, which maximizes $\norm{\partial p_\nn/\partial \mu_\nn}$.  It also leaves a big margin (close to $\pi$) between $\mu_\nn$ and other phases sharing the same cosine value; this reduces the chance of  misidentification of the phase.  We can also implement this non-adaptively by running two values of $\zzz^\myp$ apart by $\pi/2$, which allows for estimating any phase to the same precision.

\subsubsection{Calibration circuits 3}
\label{subsubsec:ins2}

\begin{figure}[tp]
\label{fig:floquet_fsim_ins2}
\centering
   \includegraphics[width=0.41347518\textwidth]{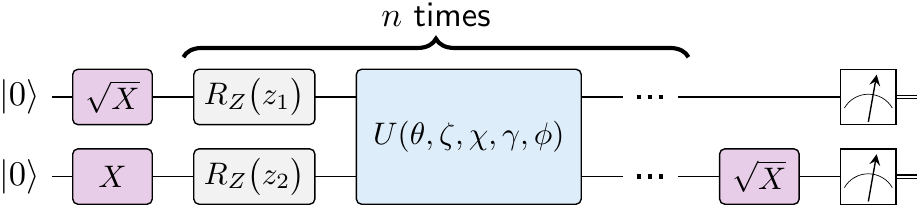}
    \caption{The circuit to calibrate 
    the parameter $\phi$. The initial microwave gates create a superposition between the state $\ket{11}$ and the single-excitation state $\ket{01}$.}
\end{figure}

This set of circuits can be used to estimate $\theta$, $\zeta$, and $\gamma+\phi$ with high precision, see Fig.~\ref{fig:floquet_fsim_ins2}.  
We will focus on $\gamma+\phi$, from which we can derive the value of the controlled phase $\phi$ given that $\gamma$ is known.  Consider the measurement probabilities
\begin{align}
  p_\nn &= \normb{\bra{10} R_{\nsp X}(0,\pi/2)\, U_c^\nn R_{\nsp X}(\pi/2,\pi)\, \ket{00}}^2\\[2pt]
  &= \frac{1}{4}\, \normb{e^{-i\nn(\gamma_c + \phi)} - \bra{1} \, u_c^\nn  \ket{0} }^2\,,\\[3pt]
   q_\nn &= \normb{\bra{00} R_{\nsp X}(0,\pi/2)\, U_c^\nn R_{\nsp X}(\pi/2,\pi)\, \ket{00}}^2\\[2pt]
   &= \frac{1}{4}\, \norm{\bra{0} \, u_c^\nn  \ket{0}}^2\,.
\end{align}
We define the relative phase
\begin{align}
  \mu_\nn &= -\nn (\gamma_c + \phi) - \arg(\bra{1} \, u_c^\nn  \ket{0})\\
& = -\nn(\gamma + \phi) + \chi + \nn\zzz^\myp + \zzz^\mym +\frac{\pi}{2}\sgn\Lambda\,,
\end{align}
which can be estimated using the same procedure as the last case.

\subsection{Phase estimation under decoherence}
\label{sec:floq_calib:phase_estimation_decoherence}

Quantum metrology protocols are typically extremely sensitive to noise~\cite{shaji_qubit_2007}.  For example, the depolarizing noise---no matter how small---ruins the possibility of sub-shot-noise performances of a quantum interferometer~\cite{ji_parameter_2008}.  It was also demonstrated that quantum metrology can only do better than classical approaches by a constant factor for any nonzero loss~\cite{kolodynski_phase_2010, knysh_scaling_2011}.  In this section, we discuss the decoherence effects on single- and two-qubit phase estimation. 

Consider a single-qubit system described by the master equation $\dot\rho=-\frac{i}{\hbar}[H,\,\rho] + \mathcal L(\rho)$, where the Lindblad operator $\mathcal L(\rho)$ takes the form
\begin{align}
\mathcal L(\rho) &=\frac{1}{T_1} \Big(\sigma^- \rho\xssp  \sigma^+ -\frac{1}{2}(\sigma^+ \sigma^-\rho +\rho  \sigma^+ \sigma^-)\Big)\nonumber\\
&\quad +
\frac{1}{2\xssp T_2} \left(\sigma^z \rho\xssp  \sigma^z - \rho \right)\,.
\end{align}
We prepare the initial state $\ket{\psi_\mathrm{in}} = (\ket{0} + \ket{1})/\sqrt 2$ and apply the phase gate $R_Z(\varphi)= \diag(1,e^{i\varphi})$ for $\nn$ times.  We then apply another phase shift $\diag(1,e^{i s})$ before measuring the system in the $X$ basis.  The probability that an excitation does not decay after $\nn$ gate cycles is $e^{-\nn \lambda_1}$, where $\lambda_1 = \text{gate time} / T_1$.  Therefore, the probability of getting the measurement outcome $+$ is
\begin{align}
p_{\nn}(s) = e^{-\nn \lambda_1} q_{\nn}(s) + \frac{1-e^{-\nn \lambda_1}}{2}\,, \label{eq:p_varphi}
\end{align}
where $q_{\nn}(s)$ is the probability of getting the outcome $+$ with only $T_2$ error
\begin{align}
 q_{\nn}(s) = \frac{1 + e^{-\nn\lambda_2}\cos(\nn\xssp \varphi + s)}{2}\,, \label{eq:q_varphi}
\end{align}
where $\lambda_2 = \text{gate time} / T_2$.  An unbiased estimator of the probability $p_{\nn}$ is
\begin{align}
   \hat p_{\nn} =  \frac{\text{Number of outcome +}}{M_n}\,,
\end{align}
where $\MM_\nn$ is the circuit repetition number.  The variance of the estimator $\hat p_\nn$ is
\begin{align}
V(\hat p_{\nn}) &=  \frac{p_{\nn}(1-p_\nn)}{\MM_\nn}\\
&=  \frac{e^{-2\nn \lambda_1} q_{\nn}(1-q_\nn)}{\MM_\nn} + \frac{1-e^{-2\nn \lambda_1}}{4 \MM_\nn}\,,
\end{align}
and the variance of the estimator of $\varphi$ can be determined using the chain rule
\begin{align}
V(\hat\varphi_\nn) &= \frac{V(\hat p_{\nn}) }{(\partial p_\nn/\partial q_\nn)^2\, (\partial q_\nn/\partial\varphi)^2}\\[3pt]
&= \frac{e^{2\nn \lambda_1}-e^{-2\nn\lambda_2}\cos(\nn\xssp \varphi+ s)^2}{\MM_\nn\xssp\nn^ 2\xssp  e^{-2\nn \lambda_2}\xssp \sin(\nn\xssp \varphi+ s)^2}\\
&\leq
\frac{e^{2\nn (\lambda_1+\lambda_2)}}{\MM_\nn\xssp\nn^ 2\xssp \sin(\nn\xssp \varphi+ s)^2}
\,.\label{eq:single_variance}
\end{align}
Equation~\eqref{eq:single_variance} diverges when $\sin(\nn\xssp \varphi + s)=0$, which can be avoided by adjusting the phase shift $s$ adaptively.  One can also implement this non-adaptively by running two experiments at $s = 0$ and $s = \pi/2$.  Using the combined information of the two, we have
\begin{align}
V(\hat\varphi_\nn) \leq  \frac{e^{2\nn (\lambda_1+ \lambda_2)}}{\MM_\nn\xssp\nn^ 2} \,.
\end{align}
Therefore, the standard deviation of the estimator $\hat\varphi_\nn$ decreases as $\nn^{-1}$ before it blows exponentially.  By setting $\partial V(\hat\varphi_\nn)/\partial n = 0$, we have
\begin{align}\label{eq:min_variance}
    \nn_\star = \frac{1}{\lambda_1 + \lambda_2}\gg 1\,,\quad V(\hat\varphi_{\nn_\star}) \leq  \frac{e^{2}(\lambda_1 + \lambda_2)^2}{ \MM_{\nn_\star}}\,.
\end{align}
The minimum variance that one can achieve is therefore set by $\lambda_1+\lambda_2$.  If the estimator $\hat\varphi_\nn$ is unbiased, the minimum variance that one can achieve is bounded by
\begin{align}\label{eq:total_variance_fisher}
V(\hat\varphi) \geq \bigg(\sum_{\nn \in \mathbb N} \frac{1}{V(\hat\varphi_\nn)}\bigg)^{-1} = \frac{1}{F}\,,
\end{align}
where $F$ is the Fisher information and $\mathbb N$ is the set of the cycle repetition numbers.  The equal sign in Eq.~\eqref{eq:total_variance_fisher} is achieved by using
\begin{align}
  \hat\varphi = \frac{1}{F}\,\sum_{\nn \in \mathbb N} \frac{\hat\varphi_\nn}{V(\hat\varphi_\nn)}\,.
\end{align}

For the two-qubit case, the impact of decoherence usually depends on the specific shapes of the control pulses.  For simplicity, we consider the resonant case $\Omega_c = \theta$ (or equivalently $\zeta_c=0$) of the two-qubit number conserving gate, where the decoherence effects are pulse-shape independent.  After removing the $T_1$ error using postselection, the measurement probability in Eq.~\eqref{eq:est_Omega} reads
\begin{align}
 p_{\nn} = \frac{1 - e^{-2\nn\lambda_2}\cos(2\nn\xssp \theta)}{2}\,, \label{eq:p_theta}
\end{align}
where the exponential factor comes from dephasing of the two qubits.  The variance of the estimator $\hat p_\nn$ is
\begin{align}
V(\hat p_{\nn}) =  \frac{p_{\nn}(1-p_\nn)}{\MM_\nn\xssp e^{-\nn \lambda_1}} \,,
\end{align}
where $\MM_n\xssp e^{-\nn \lambda_1}$ is the number of measurement pass the post selection.  The variance of the estimator of $\theta$ can be calculated by the chain rule
\begin{align}
V(\hat\theta_\nn) &= \frac{V(\hat p_{\nn}) }{(\partial p_\nn/\partial\theta)^2}\\[3pt]
&= \frac{1-e^{-4\nn\lambda_2}\cos(2\nn\xssp \theta)^2}{4\MM_\nn\xssp\nn^ 2\xssp  e^{-\nn(\lambda_1+ 4\lambda_2)}\xssp \sin(2\nn\xssp \theta)^2}\\
&\leq
\frac{e^{\nn (\lambda_1+4\lambda_2)}}{4\MM_\nn\xssp\nn^ 2\xssp \sin(2\nn\xssp \theta)^2}
\,.
\end{align}
Compared to the single-qubit case~\eqref{eq:single_variance}, the effect of $\lambda_1$ is reduced by a factor of two due to post selection while the effect of $\lambda_2$ is doubled due to the dephasing from two qubits.  By setting $\partial V(\hat\theta_\nn)/\partial n = 0$, we have
\begin{align}
    \nn_\star = \frac{2}{\lambda_1+4\lambda_2}\gg 1\,,\quad V(\hat\theta_{\nn_\star}) \leq  \frac{e^{2}\xssp(\lambda_1+4\lambda_2)^2}{4 m}\,.
\end{align}

\subsection{Cycle repetition numbers}
\label{sec:floq_clib:cyc_rep_num}

When the SPAM errors are large, the estimates from prior runs with smaller cycle repetition numbers can fail to locate the principal region of the true value.  As a result, one gets estimates with higher and higher resolutions, but in a completely wrong region.  Here we briefly discuss how to overcome this issue by properly choosing the set of cycle repetition numbers $\mathbb N$.  Consider the cost function to estimate a single-qubit phase $\varphi$,
\begin{align}
 C_\nn(x) 
 &= \sum_{\substack{\mm\in \mathbb N},\, \mm \leq \nn} \normb{e^{i\mm\xssp x}-\widehat{e^{i\mm \varphi}}}^2\;,\label{eq:cost_function}
\end{align}
where $\widehat{e^{i\mm \varphi}}$ is the estimate of $e^{i\mm \varphi}$ using quantum circuits with cycle repetition number $\mm$. 
\begin{figure}[tp]
\label{fig:cost_func}
\centering
    \includegraphics[width=0.48\textwidth]{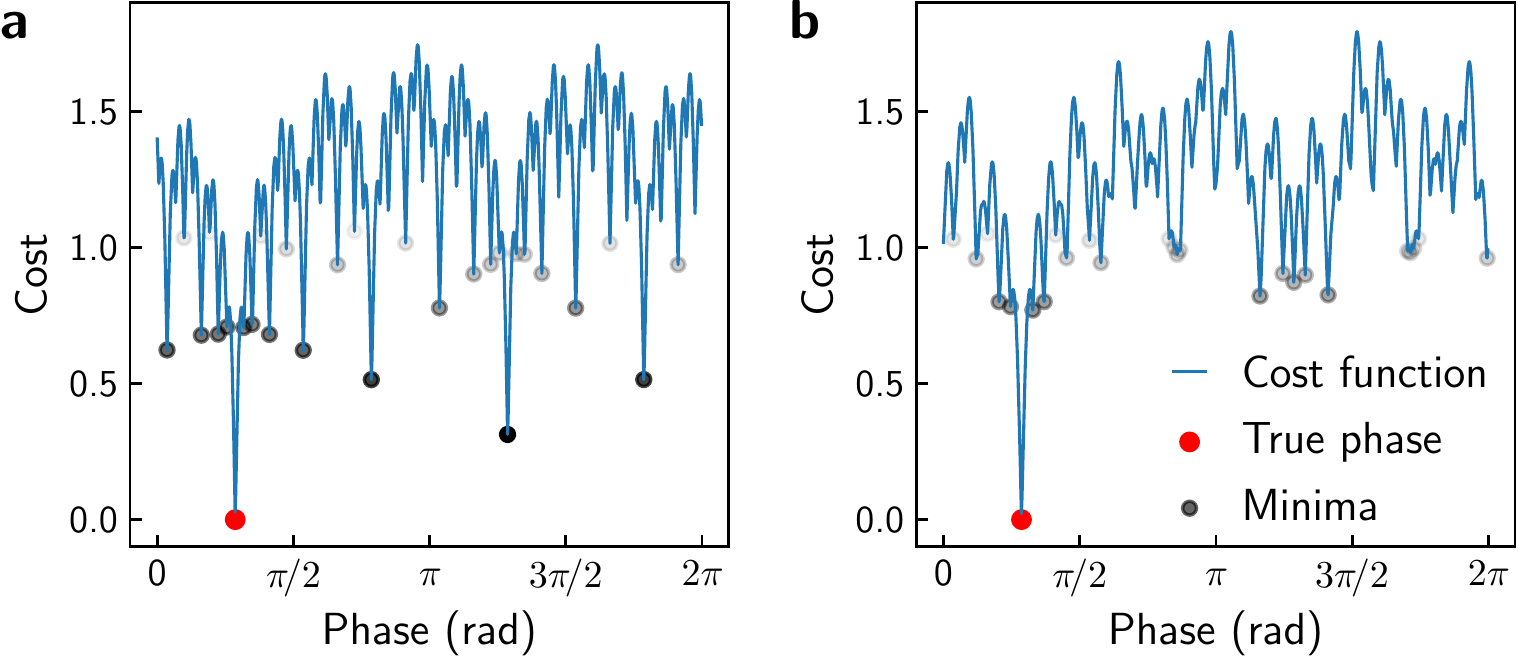}
    \caption{Comparison of two ideal cost functions using exponential circuit repetition numbers $r^k$. Left: exponent $r=2$ with $7$ runs, i.e., $\nn = 1, 2, 4, 8, 16, 32, 64$. Right: exponent $r=1.9$ with $7$ runs, i.e., $\nn = 1, 2, 4, 7, 14, 25, 48$.  There are many deep local minima that are close in value to the global minimum (red vertical line) in the left plot for $r=2$.  In comparison, the local minima are much shallower in the right plot for $r=1.9$.  This is partly due to the less spaced repetition numbers in $r=1.9$.  More importantly, it is due to the fact that the repetition numbers for $r=2$ are not mutually prime, which allows for minimizing a large number of terms in the cost function simultaneously.}
\end{figure}
A good choice of $\mathbb N$ leads to a cost function with a dominant global minima around $\varphi$.  This reduces the probability of mistaking one of the local minimum of $C_\nn(x)$ as its global minimum, which corresponds to misidentifying the principal region of the phase.  Depending on the error rates and the circuit repetition numbers, one may choose $\mathbb N = \{\lceil r^{k} \rceil \mid k = 0,1,\ldots, K-1\}$ with $r>0$.  In Fig.~\ref{fig:cost_func}, we plot the ideal cost functions by replacing $\widehat{e^{i\mm \varphi}}$ with $e^{i\mm \varphi}$ in Eq.~\eqref{eq:cost_function}.  We found that the local minima in the right panel ($r=1.9$) are much shallower than the left panel ($r=2$).  This is mainly because the numbers of cycles for $r=2$ are not mutually prime, where a large number of terms in the cost function can be minimized simultaneously.  We found that the local minima of the cost functions with larger exponents $2<r<3$ are typically shallower than those of $r=2$.  In general, we should avoid using integer $r$ and perturb the elements in $\mathbb N$ so that they are mutually prime.

\subsection{General multi-qubit unitaries}
\label{sec:floq_calib:general}

In this section, we provide a simple argument that any $\LL$-qubit gates $U$ can be determined by knowing the eigenvalues of the composite gate of the form  
\begin{align}
    U_c = U R\,,\quad R = \bigotimes_{\ell=1}^\LL R_\ell\,,
\end{align}
where $R_\ell$ acts only on the $\ell$-th qubit.  We show that $U$ can be uniquely determined by the eigenvalues of $U_c$ for various $R$.  We formally write down the eigenvalue equation $U_c\, \ket{\psi_j} = e^{-i\varepsilon_j} \ket{\psi_j}$,
where $j = 1, \ldots, 2^\LL$ and $\varepsilon_j$ is the $j$-th quasi energy.  The probability of measuring the state $\ket{\psi_\mathrm{out}}$ after applying $U_c$ to the input state $\ket{\psi_\mathrm{in}}$ for $\nn$ times is
\begin{align}
    p_{\nn} &= \bigg\lvert\sum_{j=1}^{2^\LL} e^{-i\nn\varepsilon_j} \braket{\psi_\mathrm{out}}{\psi_j}\braket{\psi_j}{\psi_\mathrm{in}}
    \bigg\rvert^2\\
    &=\sum_{j,k=1}^{2^\LL} e^{-i\nn(\varepsilon_j-\varepsilon_k)} a_j a_k^*\,,
\end{align}
where $a_j = \braket{\psi_\mathrm{out}}{\psi_j}\braket{\psi_j}{\psi_\mathrm{in}}$.  In principal, the difference $\varepsilon_j- \varepsilon_k$ can be estimated by running the circuits with different $\nn$ and $\ket{\psi_\mathrm{in}}$ and $\ket{\psi_\mathrm{out}}$.      

To determine $U$, it suffices to learn $\tr (U R)$ for a set of product unitaries $R$ that form a complete operator basis, e.g., the Pauli group $\mathcal P$,
\begin{align}
   U = \frac{1}{2^\LL}\, \sum_{R\in\mathcal P}  \tr (U R) \, R\,.
\end{align}
Without losing generality, we assume that $\det U=\det R = \det U_c = 1$; therefore, we have the condition $\sum_j \varepsilon_j = 0 \; (\mathord{\bmod}\; 2\pi)$.  Knowing the differences in the quasi energies allows one to determine $\varepsilon_j \; (\mathord{\bmod}\; 2\pi/2^{\LL})$.  The trace
$\tr U_c = \tr (U R) = \sum_j e^{-i\varepsilon_j}$ can therefore be estimated up to a phase factor $e^{ik \pi/2^{\LL-1}}$, where $k$ is an integer.  To get the correct phase factor, we introduce a continuous set of single-qubit unitaries
\begin{align}
    R(s) = \bigotimes_{\ell=1}^\LL e^{is\bm\sigma\cdot \bm v_\ell}\,,
\end{align}
where $\bm{\sigma} = (\sigma^x\; \sigma^y\; \sigma^z)$ is the vector of Pauli operators and the normalized vector $\bm v_\ell$ determines the rotational axes of $R_\ell$.  By choosing a sequence values $0\leq s \leq 1$, one can keep track the principal region of the overall phase of $\tr (UR)$.  The procedure described here is by no means optimal, but it shows that our method can be generalized to multi-qubit gates in principle.

\putbib[fhm]
\end{bibunit}

\clearpage

\begin{figure*}
\centering
\begin{minipage}[t]{.48\textwidth}\vspace{0pt}
\includegraphics[width=0.95\textwidth]{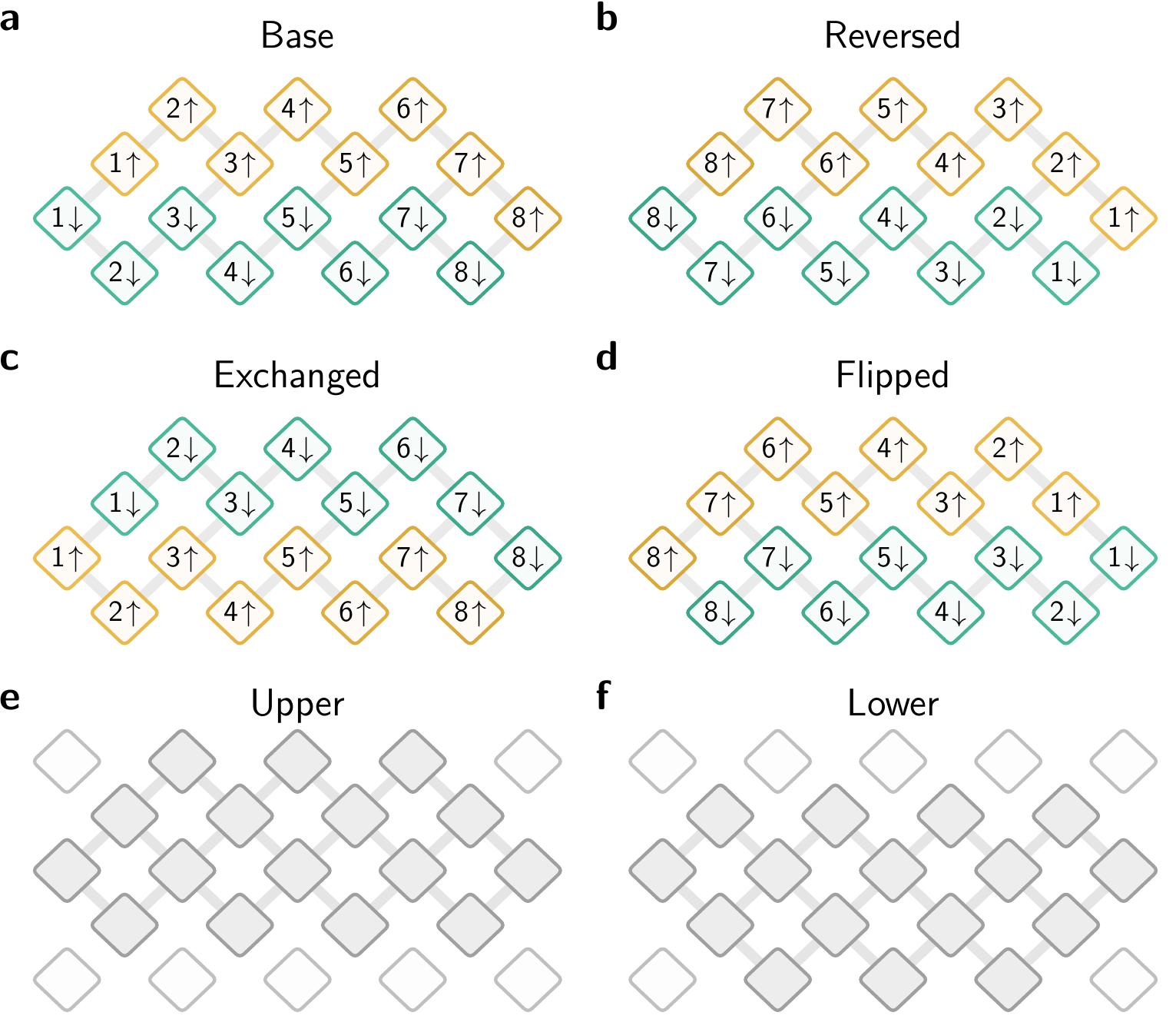}
\caption{Various qubit assignments of the 1D Fermi-Hubbard model on a 23-qubit grid. \textbf{a.} The original configuration is used as the base.  We generate new configurations by applying the following operations and their combinations on the base: \textbf{b.} reversing the sites, \textbf{c.} exchanging the spin states, \textbf{d.} flipping the sites horizontally.  Each configuration can either take one of the two subsets of the grid: \textbf{e.} the upper part, \textbf{f.} the lower part.  This leads to $16$ different qubit assignments.  The same assignments are used regardless if interaction terms are present ($U \neq 0$) or not ($U = 0$).} 
\label{fig:different_layouts}
\end{minipage}\hfill
\begin{minipage}[t]{.48\textwidth}\vspace{0pt}
\centering
  \includegraphics[width=0.96\textwidth]{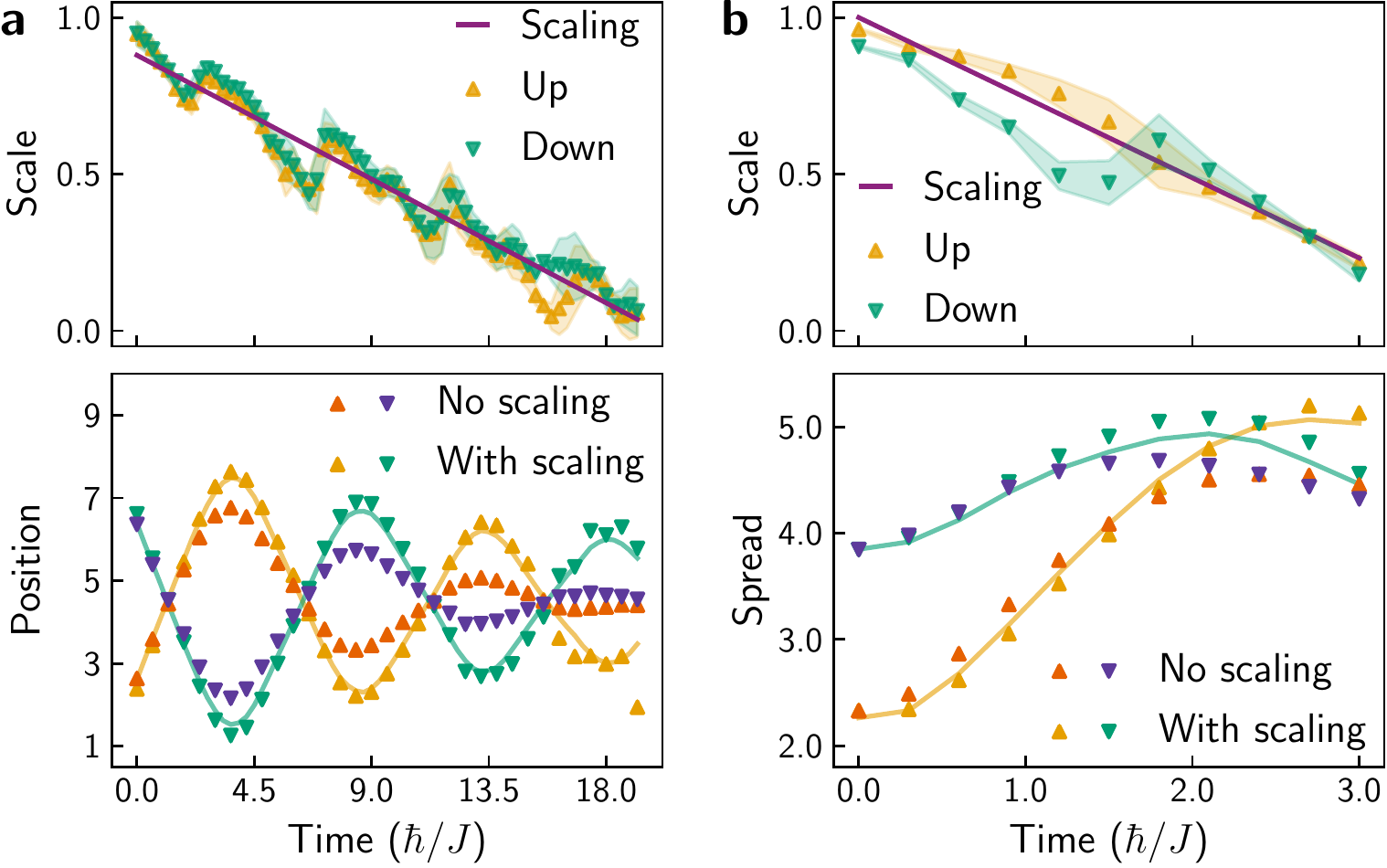}
    \caption{The damping factor (scale): \textbf{a.} the $U=0$ and $N_P=2$ case in Fig.~\ref{fig:two_gaussians}, \textbf{b.} the $U=2$ and $N_P=4$ case in Fig.~\ref{fig:scs}.  The data points are obtained by comparing the numerical and experimental results at each Trotter step.  The shaded areas represent the standard errors of regression and the solid lines represent the linear fittings of the data points.  The bottom plots show the effectiveness of the rescaling procedure using the linear relation in Eq.~\eqref{eq:rescale}: \textbf{a.} center of mass positions of the spin-up (yellow and orange) and spin-down (green and purple) states, \textbf{b.} spreads of the two spin states.}
    \label{fig:scaling}
\end{minipage}
\end{figure*}

\begin{figure*}
\centering
\begin{minipage}[t]{.48\textwidth}\vspace{0pt}
\includegraphics[width=0.9\textwidth]{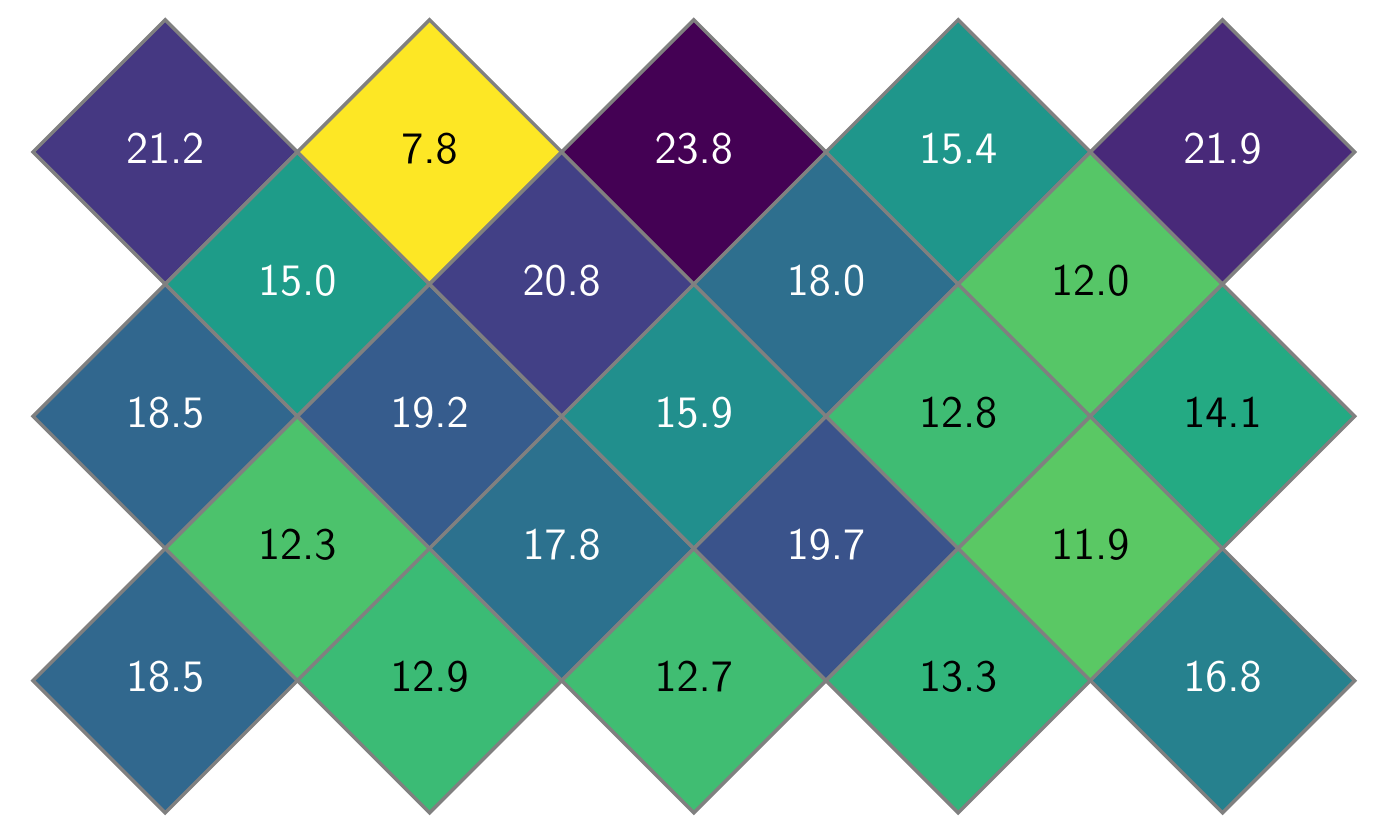}
\caption{$T_1$ relaxation times (\SI{}{\micro\second}) at idle frequencies.} 
\label{fig:t1}
\end{minipage}\hfill
\begin{minipage}[t]{.48\textwidth}\vspace{0pt}
\centering
  \includegraphics[width=0.9\textwidth]{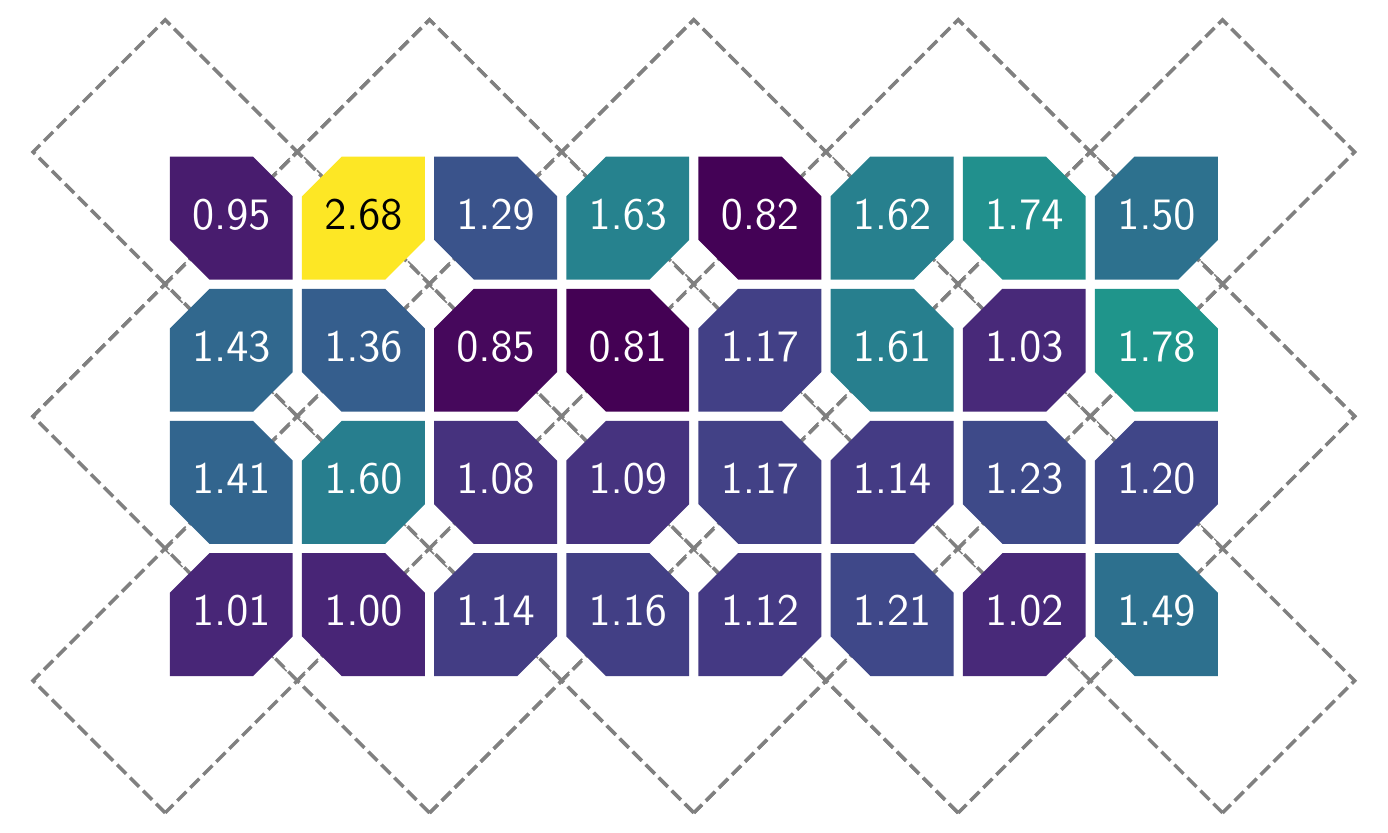}
    \caption{Two-qubit gate percent Pauli error obtained by cross-entropy benchmarking (XEB), where non-conflicting subsets of two-qubit gates are applied simultaneously.  These values were measured against the standard $\sqrt{\iSWAP}^{\,\dag}$ without the parasitic controlled phases.  The two-qubit gates are calibrated using our routine calibration process without Floquet calibration, see Section V in the Supplementary information in~\cite{arute_quantum_2019}.}
    \label{fig:xeb}
\end{minipage}
\end{figure*}

\begin{figure*}[ht]
   \centering
   \subfloat{\includegraphics[width=0.99\textwidth]{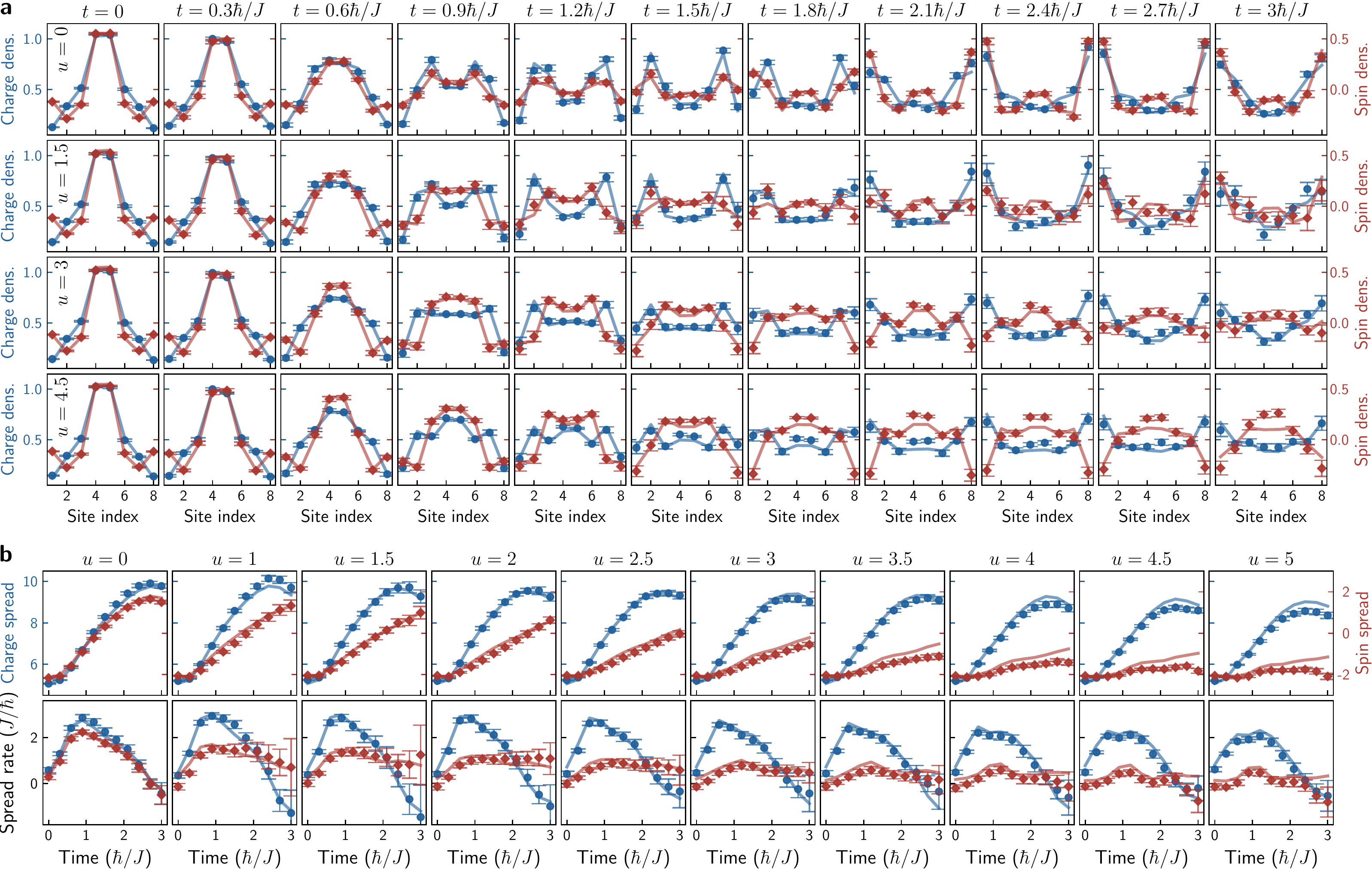}}
    \caption{Detailed data for the $N_\uparrow=N_\downarrow=2$ case. \textbf{a.} Charge and spin densities $\rho_j^\pm$ as functions of the site numbers at different evolution times for interaction strengths $u = 0, 1.5, 3, 4.5$. \textbf{b.} Charge and spin spread $\kappa^\pm$ and their derivatives as functions of time for different values of $u$.}
   \label{fig:full_data_2_2} 
\end{figure*}

\begin{figure*}[ht]
   \centering
   \subfloat{\includegraphics[width=0.99\textwidth]{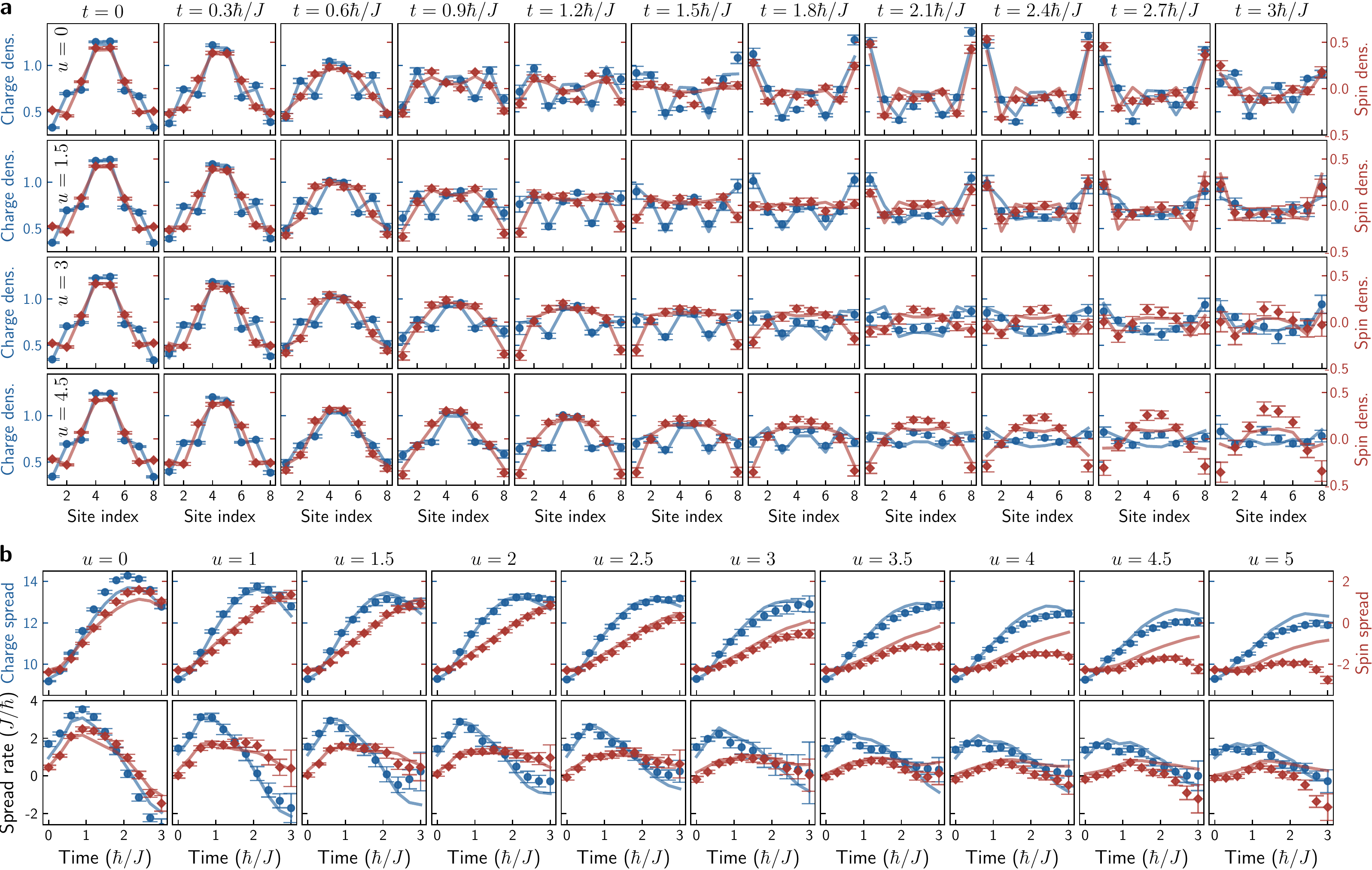}}
   \caption{Detailed data for the $N_\uparrow=N_\downarrow=3$ case. \textbf{a.} Charge and spin densities $\rho_j^\pm$ as functions of the site numbers at different evolution times for interaction strengths $u = 0, 1.5, 3, 4.5$. \textbf{b.} Charge and spin spread $\kappa^\pm$ and their derivatives for different values of $u$.}
   \label{fig:full_data_3_3}
\end{figure*}

\end{appendices}

\end{document}